\documentclass[12pt, a4paper]{article}
\usepackage{amsfonts, amssymb}
\usepackage{graphicx}
\usepackage{latexsym,amsmath,color,array}

\usepackage{natbib}
\usepackage{enumerate}
\usepackage{bbding}
\usepackage{amssymb}
\usepackage{amsmath}
\usepackage{graphicx}
\usepackage{amsmath}
\usepackage{bbm}
\usepackage{mathtools}
 \usepackage{color}
 \usepackage{array}
 \usepackage{multirow}
 \usepackage[dvipsnames]{xcolor}
 \usepackage{makecell}
 \usepackage{colortbl}
 \usepackage[flushleft]{threeparttable} % http://ctan.org/pkg/threeparttable
\usepackage{rotating} % To display tables in landscape
\usepackage{enumitem}
\usepackage{subfig}
\usepackage{adjustbox}
\usepackage{xcolor}
\usepackage{bm}
\def\1{\mathbb{I}}

\newcounter{thm}[section]
\newcounter{appen}[section]
\newcounter{assum}[section]

\begin{document}

\title{Penalized Variable Selection in Multi-Parameter Regression Survival Modelling}
\author{Fatima-Zahra Jaouimaa\footnote{Department of Mathematics and Statistics, University of Limerick; fatima.jaouimaa@ul.ie} \hspace{3cm}
Il Do Ha\footnote{Department of Statistics, Pukyong National University; idha1353@pknu.ac.kr} \hspace{3cm}
Kevin Burke\footnote{Department of Mathematics and Statistics, University of Limerick; kevin.burke@ul.ie}  }
\date{\today}

\maketitle

\begin{abstract}
Multi-parameter regression (MPR) modelling refers to the approach whereby covariates are allowed to enter the model through multiple distributional parameters simultaneously. This is in contrast to the standard approaches where covariates enter through a single parameter (e.g., a location parameter). Penalized variable selection has received a significant amount of attention in recent years: methods such as the least absolute shrinkage and selection operator (LASSO), smoothly clipped absolute deviation (SCAD), and adaptive LASSO are used to simultaneously select variables and estimate their regression coefficients. Therefore, in this paper, we develop penalized multi-parameter regression methods and investigate their associated performance through simulation studies and real data; as an example, we consider the Weibull MPR model.
\smallskip

{\bf Keywords.} MPR; Weibull; Penalized maximum likelihood; Differential  evolution  algorithm.

\end{abstract}

\qquad

\newpage
\section{Introduction}
\label{sec:intro}
		Multi-parameter regression (MPR) modelling refers to the approach whereby covariates are allowed to enter the model through multiple distributional parameters simultaneously. This is in contrast to the standard approaches where covariates enter through a single parameter (e.g., a location parameter) while holding the remaining parameter(s) (e.g., a scale or shape parameters) constant. Multi-parameter regression approaches have been considered in many areas of applied statistics. One of the earliest examples appears in econometrics literature where \citet{park1966estimation} proposed a log-linear model for the scale parameter in the normal linear model when the assumption of homoscedasticity is violated and describes a two stage process for the estimation of the parameters. \citet{harvey1976estimating} develops a maximum likelihood estimation procedure for these so called heteroscedastic regression models. \citet{stirling1985heteroscedastic} and \citet{aitkin1987modelling} illustrate the procedure using examples and GLIM (Generalized Linear Interactive Modelling) macros for the implementation of these models. Other examples of multi-parameter regression models include an extension of generalized linear models to allow for the joint modelling of the mean and dispersion \citep{smyth1989generalized, nelder1991generalized, lee1998generalized}, the generalized additive models for location, scale and shape (GAMLSS) which goes beyond the exponential family to general parameters (not necessarily location and scale) for a variety of distributions \citep{rigby2005generalized}. \citet{taylor2004joint} model the location and scale parameters of the t-distribution jointly to rectify the dependence of the scale parameter on the location. %\citet{wu2012variable} propose a joint mean and dispersion model of the inverse Gaussian distribution and argue that an efficient estimation of the mean parameters in regression is dependent on the correct the modelling of the dispersion parameter. To model asymmetric and heavy tailed data \citet{wu2013variable} consider a joint location-scale skew-normal model, and this was generalized to a skew-t-normal model by \citet{wu2014variable}.
		\par
		In this article, we consider MPR modelling in the setting of survival analysis. Examples of the MPR modelling framework include \citet{anderson1991nonproportional} who extended the Weibull accelerated failure time model such that both the location and dispersion parameters depend on covariates. A set of models known as the threshold regression models, model survival time through an underlying diffusion process (e.g., Weiner process) whose drift and initial state parameters both depend on covariates \citep{lee2006threshold, aalen2008survival}. More recently, \citet{burke2017multi} explored the general use of parameteric hazard MPR models, and demonstrated the favourable interpretation and flexibility afforded by jointly modelling the scale and shape parameters of a Weibull distribution.
		\par
		The variable selection problem is omnipresent in most, if not all, statistical applications. This problem arises due to the uncertainty associated with the selection of a subset of explanatory variables to model some variable of interest \citep{thompson1978selection, george2000variable}. With the growing availability of data, this problem has received a significant amount of attention in recent years. Traditionally, variable selection has been performed using methods such as backward elimination, forward selection, and stepwise regression; \citet{miller2002subset} provides a comprehensive discussion of these methods. Due to their inherent discreteness {(i.e., covariates are either ``in'' or ``out'')}, these methods can be unstable \citep{breiman1996heuristics}. Furthermore, they are known to be computationally intensive with a power explosion of the number of possible submodels ($2^{p}$) to be considered for $p$ predictors. Modern approaches have been focused on penalized regression whereby parameter estimation and (continuous) model selection are both carried simultaneously. Methods such as the least absolute shrinkage and selection operator (LASSO) \citep{tibshirani1996regression}, smoothly clipped absolute deviation (SCAD) \citep{fan2001variable}, the elastic net \citep{zou2005regularization}, and adaptive lasso (ALASSO) \citep{zou2006adaptive} are used to simultaneously select variables and estimate their regression coefficients.
		\par
		Standard regression approaches only have one regression component, and, therefore, variable selection literature mainly focuses on this, i.e., selection of covariates in a location parameter (cf.~\citet{fan2010selective}). Beyond this standard setting, MPR models require variable selection in multiple distributional parameters. However, little work has been done in this area. \citet{wu2012variable} considered penalized likelihood for variable selection in the case of an inverse Gaussian joint mean and dispersion model, while \citet{wu2013variable} and \citet{wu2014variable} applied this procedure to joint location-scale skew-normal and skew-t-normal models. None of the aforementioned papers consider censored data, and, furthermore, these papers only consider the case of a common tuning parameter for the two regression components. Therefore, the main objective of this article is to develop penalized variable selection in MPR models more generally. Using the Weibull MPR model as an example, we investigate the need for a separate tuning parameter for each regression component.  We select tuning parameters based on the BIC function, and, because this is multi-modal, we propose the use of a differential evolution ``global'' optimization procedure.
		\par
		The remainder of the paper is structured as follows. In Section 2, the Weibull multi-parameter regression model, the penalty functions and the penalized likelihood estimation procedure are introduced. Section 3 describes the model fitting process and the algorithm used in the selection of the tuning parameter(s). Simulation studies to investigate the characteristics of the algorithm and evaluate its performance in both variable selection and parameter estimation are given in Section 4. The proposed methodologies are then illustrated on a real data example, a lung cancer dataset, in Section 5. A discussion of the proposed methods and some concluding remarks are given in the last section, Section 6.
%\section{Penalized Estimation in the Weibull MPR Model}
\section{Model Formulation}
\label{sec:Formulation}
 \subsection{Weibull MPR Model}
		Although the variable selection methods we consider in this article can be applied to any parametric MPR model, it is helpful to focus on a specific example. We therefore consider the Weibull MPR model as the Weibull distribution is one of the most popular parametric survival distributions. {The Weibull hazard function for survival time $\tilde{T}_i$ corresponding to the $i$th individual is given by }
		\begin{equation}\nonumber
		h(t|x_i,z_i)=\tau_i\gamma_i t^{\gamma_i-1} \ ,
		\end{equation}
		for $i = 1, 2, \dots, n$, where $\tau_i>0$ is the scale parameter and $\gamma_i>0$ is the shape parameter. The Weibull MPR model is obtained by letting both distributional parameters depend on covariates as follows:
		$$\log(\tau_i)=x^{T}_{i}\beta, \qquad \log(\gamma_i)=z^{T}_{i}\alpha,$$
		where $x_i=(1,x_{i1},\dots,x_{ip})^T$ and $z_i=(1,z_{i1},\dots,z_{iq})^T$ are scale and shape covariate vectors which may or may not have covariates in common, $\beta=(\beta_0,\beta_1,\dots,\beta_p)^T$ and $\alpha=(\alpha_0,\alpha_1,\dots,\alpha_q)^T$ are the corresponding regression coefficients, and the log link is used to ensure positivity of the parameters.
		\par
		Parameter estimation within the unpenalized MPR model can be carried out in a standard fashion using maximum likelihood. {First, let $T_i = \text{min}(\tilde{T}_i, C_i) $ be the observed survival time for the $i$th individual. Then the associated log-likelihood function given by  }
	 	\begin{equation}\label{eq:NonPen_lik}
		\ell(\theta)= \sum^{n}_{i=1}{\delta_i \{\log{\tau_i} + \log{\gamma_i} + (\gamma_i-1) \log{t_i}\}}- \tau_i t^{\gamma_i}_i,
		\end{equation}
		where $\theta=(\beta^T,\alpha^T)^T$ is the full parameter vector, {$t_i$ is the realization of $T_i$}, and $\delta_i$ is the censoring indicator which takes the value 0 for censored survival times and 1 for uncensored survival times. Beyond the Weibull case we consider here, the likelihood function is $\sum_{i=1}^n \delta_i \log h(t_i|x_i,z_i) - H(t_i|x_i,z_i)$ where $H(t|x_i,z_i) = \int_0^t h(u|x_i,z_i) du$ is the cumulative hazard function.

	\subsection{Penalized Likelihood\label{sec:penalizedlike}}
		Penalized MPR estimation can be developed on the basis of maximising a penalized log-likelihood given by
		\begin{equation}\label{eq:Model1}
		\ell_\lambda(\theta)= \ell(\theta)- n \sum^{p}_{j = 0} J_{\lambda_{\beta_j}}(\lvert\beta_j\rvert) - n \sum^{q}_{j = 0} J_{\lambda_{\alpha_j}}(\lvert\alpha_j\rvert),
		\end{equation}
		where $\ell(\theta)$ is the unpenalized likelihood, $\lambda = (\lambda_{\beta_0}, \lambda_{\beta_1}, \ldots,\lambda_{\beta_p}, \lambda_{\alpha_0}, \lambda_{\alpha_1}, \ldots,\lambda_{\alpha_q})$ is a vector of coefficient-specific tuning parameters, and $J_{\lambda_{\beta_j}}(\cdot)$ and $J_{\lambda_{\alpha_j}}(\cdot)$ are scale and shape penalty functions which we assume have the same functional form (but differ with respect to the tuning parameter). As is standard practice, we assume that the intercepts are not penalized, and, therefore, define $\lambda_{\beta_0} \equiv \lambda_{\alpha_0} \equiv 0$ (rather than, for example, assuming the intercepts are zero as other authors do); we also assume that the covariates are standardized. Although we have defined $\lambda$ quite generally, we will in fact impose constraints on this vector (beyond fixing $\lambda_{\beta_0} \equiv \lambda_{\alpha_0} \equiv 0$) by considering the following possibilities (for $j \ne 0$):
 \begin{enumerate}[label=\roman*)]
   \item single penalty,
   $$\lambda_{\beta_j} = \lambda_{\alpha_j} = \lambda,$$
   \item single adaptive penalty,
   $$\lambda_{\beta_j} = \lambda w_{\beta_j}, \qquad \lambda_{\alpha_j} = \lambda w_{\alpha_j},$$ where $w_{\beta_j}$ and $w_{\alpha_j}$ are predefined weights,
   \item separate non-adaptive penalties,
   $$\lambda_{\beta_j} = \lambda_{\beta}, \qquad \lambda_{\alpha_j} = \lambda_\alpha,$$
   \item separate adaptive penalties,
   $$\lambda_{\beta_j} = \lambda_{\beta} w_{\beta_j}, \qquad \lambda_{\alpha_j} = \lambda_{\alpha} w_{\alpha_j}.$$
		\end{enumerate}
		Note that the only ``adaptive" penalty considered for the purpose of this article is the ALASSO penalty. (i) and (ii) are standard approaches where a single penalty, $\lambda$, applies to the whole vector of parameters. This is reasonable in more standard setting where there is only a $\beta$ vector.  However, in this particular MPR setting, we have two separate distributional parameters, which exist on different scales. For this reason we investigate methods (iii) and (iv) which apply different penalties to the two regression vectors via $\lambda_\beta$ and $\lambda_\alpha$.
		\par
		For the purpose of this article, we consider the most commonly used penalties, namely: the least absolute shrinkage and selection operator (LASSO) \citep{tibshirani1996regression},
		\begin{equation*}
		J_{\lambda_{\theta_j}}(\lvert\theta_j\rvert) = \lambda_{\theta_j}\lvert\theta_j\rvert,
		\end{equation*}
		which, although popular, is known to select too many variables \citep{radchenko2008variable}; the non-convex smoothly clipped absolute deviation (SCAD) \citep{fan2001variable},
		\begin{equation*}
			J_{\lambda_{\theta_j}}(\lvert\theta_j\rvert) =  \left\{
				\begin{array}{ll}
				\lambda_{\theta_j}(\lvert\theta_j\rvert) & \qquad \text{if }\ \lvert\theta_j\rvert \le  \lambda_{\theta_j},\\
				\frac{2a \lambda_{\theta_j} \lvert\theta_j\rvert - \theta_{j}^2 -  \lambda_{\theta_j}^2}{2(a-1)} & \qquad \text{if }\  \lambda_{\theta_j} <  \lvert\theta_j\rvert < a \lambda_{\theta_j},\\
				\frac{\lambda_{\theta_j}^2(a+1)}{2} & \qquad \text{if }\ \lvert\theta_j\rvert \ge a \lambda_{\theta_j},\\
				\end{array}
				\right.
				\end{equation*}
				where $a = 3.7$, and the adaptive LASSO (ALASSO) \citep{zou2006adaptive},
				\begin{equation*}
					J_{\lambda_{\theta_j}}(\lvert\theta_j\rvert) = \lambda_{\theta_j} w \lvert\theta_j\rvert\ ,
				\end{equation*}
				where, typically, {$w=1/\lvert\hat\theta_{0,j}\rvert$ and $\hat\theta_{0,j}$} is a unpenalized estimate of $\theta_j$. These so called adaptive weights are used to apply different penalties to different regression coefficients such that a larger amount of shrinkage is applied to the unimportant variables. Note that, here, we use $\theta_j$ to denote a generic regression coefficient, and $\lambda_{\theta_j}$ is the corresponding tuning parameter. The latter two penalties (i.e., SCAD and ALASSO) are known to possess the oracle property, i.e., the procedure asymptotically identifies the right subset model and estimates the coefficients and covariance matrix as though the true model were known in advance \citep{fan2001variable}.
		\section{Penalized Estimation Procedure}
		\label{sec:Estimation}
 \subsection{Model Fitting}
 We define
		\begin{equation}\label{eq:Est_theta}
			\hat{\theta}_{\lambda} = \underset{\theta}{\text{argmax}}\ \ell_{\lambda}(\theta),
		\end{equation}
		where $\ell_\lambda(\theta)$ is given by (\ref{eq:Model1}). The corresponding score functions are given by %Hence, the penalized maximum likelihood estimating equations are given by solving
     \begin{align} \label{eq:Pen_Lik}
            	\begin{split}	
            	\frac{\partial \ell_\lambda}{\partial \beta} &= \frac{\partial \ell}{\partial \beta} - nV_{\beta} = X^T U_{\beta} - nV_{\beta},\\
            	\frac{\partial \ell_\lambda}{\partial \alpha} &= \frac{\partial \ell}{\partial \alpha} - nV_{\alpha} = Z^T U_{\alpha} - nV_{\alpha},
            	\end{split}
         \end{align}
     where $X$ is an $n\times(p+1)$ matrix whose $i$th row is $x_i$, $Z$ is an $n\times(q+1)$ matrix whose $i$th row is $z_i$; $U_{\beta}$ and $U_{\alpha}$ are vectors of length $n$ such that $U_{\beta i} = \delta_i - \tau_i t^{\gamma_i}_i$ and $U_{\alpha i} = \delta_i (1 + \gamma_i \log{t_i}) - \tau_i\gamma_i t^{\gamma_i}_i \log{t_i}$; $V_\beta$ and $V_\alpha$ are vectors of lengths $p+1$ and $q+1$ respectively, such that, for $j \ge 0$, $V_{\beta,j+1} = d J_{\lambda_{\beta_j}}(\lvert\beta_j\rvert) / d \beta_j
     = J_{\lambda_{\beta_j}}'(\lvert\beta_j\rvert) d \lvert\beta_j\rvert / d\beta_j$ and $V_{\alpha, j+1} =d J_{\lambda_{\alpha_j}}(\lvert\alpha_j\rvert) / d \alpha_j = J_{\lambda_{\alpha_j}}'({\lvert\alpha_j\rvert}) d \lvert\alpha_j\rvert / d\alpha_j$.
     \par
		Note however, the presence of the absolute value function renders the penalty functions non-differentiable at zero. Various algorithms have been developed to overcome this issue including quadratic programming \citep{tibshirani1996regression}, least angle regression (LARS) \citep{efron2004least}, co-ordinate descent \citep{friedman2007pathwise} and the local quadratic approximation \citep{fan2001variable}. In this paper, we take a different approach, and use an extension of the absolute value function given by {$$a(x) = \sqrt{x^2+\epsilon^2}-\epsilon,$$ where $\lim_{\epsilon \to 0} a(x) = \lvert x \rvert $. This yields a differentiable penalty so that standard gradient-based optimization algorithms can be applied straightforwardly and transparently.} Thus, $a'(x) = x/ \sqrt{\epsilon^2 + x^2}$ (which is an approximation of the signum function) and $a''(x) = {\epsilon^2}/{(\epsilon^2 + x^2)^{3/2}}$. Smaller values of $\epsilon$ bring the approximate penalty closer to the original penalty, but also closer to the penalty being non-differentiable; we have found that fixing $\epsilon=10^{-4}$ generally works well. As we use smooth $J(\cdot)$ functions, and $a(x)$ in place of $|x|$, (\ref{eq:Pen_Lik}) is then smooth in the parameters and can therefore be solved using the Netwon-Raphson algorithm.

{ We denote by $I_\lambda(\theta)$ the matrix of second derivatives of $\ell_\lambda(\theta)$, i.e., $-\nabla_{\theta}\nabla_{\theta}^T\ell_\lambda(\theta)$. Then,
\begin{align*} \label{eq:Imat}%\nonumber
I_\lambda(\theta)
= I_0(\theta) +
\begin{pmatrix}
             n\Sigma_{\beta}
             & 0 \\
             0 & n\Sigma_{\alpha}
         \end{pmatrix}
=   \begin{pmatrix}
             X^{T} W_{\beta} X + n\Sigma_{\beta}
             & X^{T} W_{\alpha\beta} Z \\
             Z^{T} W_{\alpha\beta} X & Z^{T} W_{\alpha} Z + n\Sigma_{\alpha}
         \end{pmatrix}
\end{align*}
where $I_0(\theta) = -\nabla_{\theta}\nabla_{\theta}^T\ell(\theta)$ is the usual observed information matrix of the unpenalized likelihood; $\Sigma_{\beta}$ and $\Sigma_{\alpha}$ appear due to the penalties,} and are diagonal matrices of dimension $(p+1)\times(p+1)$ and $(q+1)\times(q+1)$, respectively, such that, for $j \ge 0$, $\Sigma_{\beta,j+1,j+1} = d^2 J_{\lambda_{\beta_j}}(\lvert\beta_j\rvert) / {d \beta_j}^2$ and $\Sigma_{\alpha,j+1,j+1} = d^2 J_{\lambda_{\alpha_j}}(\lvert\alpha_j\rvert) / {d \alpha_j}^2$; and $W_{\beta}$, $W_{\alpha}$, and $W_{\alpha\beta}$ are $n \times n$ diagonal matrices whose $i$th diagonal elements are given by $\tau_i t^{\gamma_i}_{i}$, $\{\tau_i t^{\gamma_i}_i (\gamma_i \log{t_i} + 1 ) -\delta_i\}\gamma_i \log{t_i}$, and $\tau_i \gamma_i t^{\gamma_i}_{i} \log{t_i}$ respectively. Thus, following \citet{ha2014variable}, the resulting system of Newton-Raphson equations, which are iteratively solved for {$\theta_\lambda^{(m+1)} = ({\beta_\lambda^{(m+1)}}^T, {\alpha_\lambda^{(m+1)}}^T)^T$}, can be written compactly as
\begin{equation} \label{eq:NR}%\nonumber
			\begin{pmatrix}
             X^{T} W_{\beta}^{(m)} X + n\Sigma_{\beta}^{(m)}
             & X^{T} W_{\alpha\beta}^{(m)} Z \\
             Z^{T} W_{\alpha\beta}^{(m)} X & Z^{T} W_{\alpha}^{(m)} Z + n\Sigma_{\alpha}^{(m)}
         \end{pmatrix}
         %\quad
         {
         \begin{pmatrix}
             \beta_\lambda^{(m+1)} - \beta_\lambda^{(m)}\\
             \alpha_\lambda^{(m+1)} - \alpha_\lambda^{(m)}
         \end{pmatrix}}
         %\quad
         =
         %\quad
         \begin{pmatrix}
             X^T U_{\beta}^{(m)} - nV_{\beta}^{(m)}\\
             Z^T U_{\alpha}^{(m)} - nV_{\alpha}^{(m)}
     \end{pmatrix},
\end{equation}
{ where the various elements super-scripted by $(m)$ depend on $\theta_\lambda^{(m)}$, but this dependence is suppressed for notational convenience; we use unpenalized estimates as the initial values in this iterative procedure, i.e., $\theta^{(0)} = \hat\theta_0$. Having obtained the penalized estimates, $\hat\theta_\lambda$, the covariance can be estimated using the sandwich formula
 		\begin{equation} \label{eq:SE_formula}
 			\hat{cov}(\hat{\theta}_{\lambda}) = \{I_\lambda(\hat{\theta}_\lambda)\}^{-1} I_0(\hat{\theta}_\lambda) \{I_\lambda(\hat{\theta}_\lambda)\}^{-1}
		 \end{equation}}
\citep{fan2001variable, fan2002variable,ha2014variable, park2018penalized}. This formula is known to have good accuracy when the sample size is moderate \citep{fan2001variable, fan2002variable}, and its performance in our MPR setting is investigated in Section \ref{sec:Simulations} through simulation studies.
		
		\subsection{Tuning Parameter Selection}
		The selection of ``optimal'' tuning parameter(s) is typically done through the use of data-driven criteria such as generalized cross-validation (GCV), Akaike information criterion (AIC) or Bayesian information criterion (BIC). GCV and the AIC are known to be asymptotically ``loss-efficient'' and ``selection inconsistent'' variable selection criteria \citep{shao1997asymptotic, yang2005can, wang2009shrinkage}. \citet{wang2007tuning} provide a formal proof that the shrinkage or tuning parameter selected using GCV may not be able to identify the true model consistently for the SCAD estimator in linear models and partially linear models. Instead, they suggest using the BIC and prove its model selection consistency property. A similar conclusion has been reached by \citet{wang2007unified} for the ALASSO. Hence, due to its widely reported superior empirical performance in variable selection, we use a BIC-type criterion to determine the values of the tuning parameter(s), where
		% i.e. the tuning parameter selected by the BIC can identify the true model with a probability tending to 1.
		\begin{equation}
		\text{BIC}(\lambda)= -2 \ell(\hat{\theta}_{\lambda}) + e_{\lambda} \log{n},
		\label{eq:BIC}
		\end{equation}
		$\ell(\hat{\theta}_\lambda)$ is the unpenalized likelihood function defined in (\ref{eq:NonPen_lik}), $n$ is the sample size and {$e_\lambda = \text{tr}[\{I_\lambda(\hat{\theta}_\lambda)\}^{-1} I_0(\hat{\theta}_\lambda)]$} is the effective degrees of freedom \citep{ha2007model}; we define
		\begin{equation}
\lambda^{*} = \underset{\lambda}{\text{argmin}}\ \text{BIC}(\lambda).
		\label{eq:lambdastr}
		\end{equation}
{ Note that, as described in Section \ref{sec:penalizedlike}, $\lambda^*$ will either be one-dimensional (when a common penalty is applied to $\beta$ and $\alpha$) or two-dimensional (when separate penalties are applied).}

		\par
\begin{figure}[!ht]
			\begin{center}
             \subfloat[]{{\includegraphics[scale=0.6]{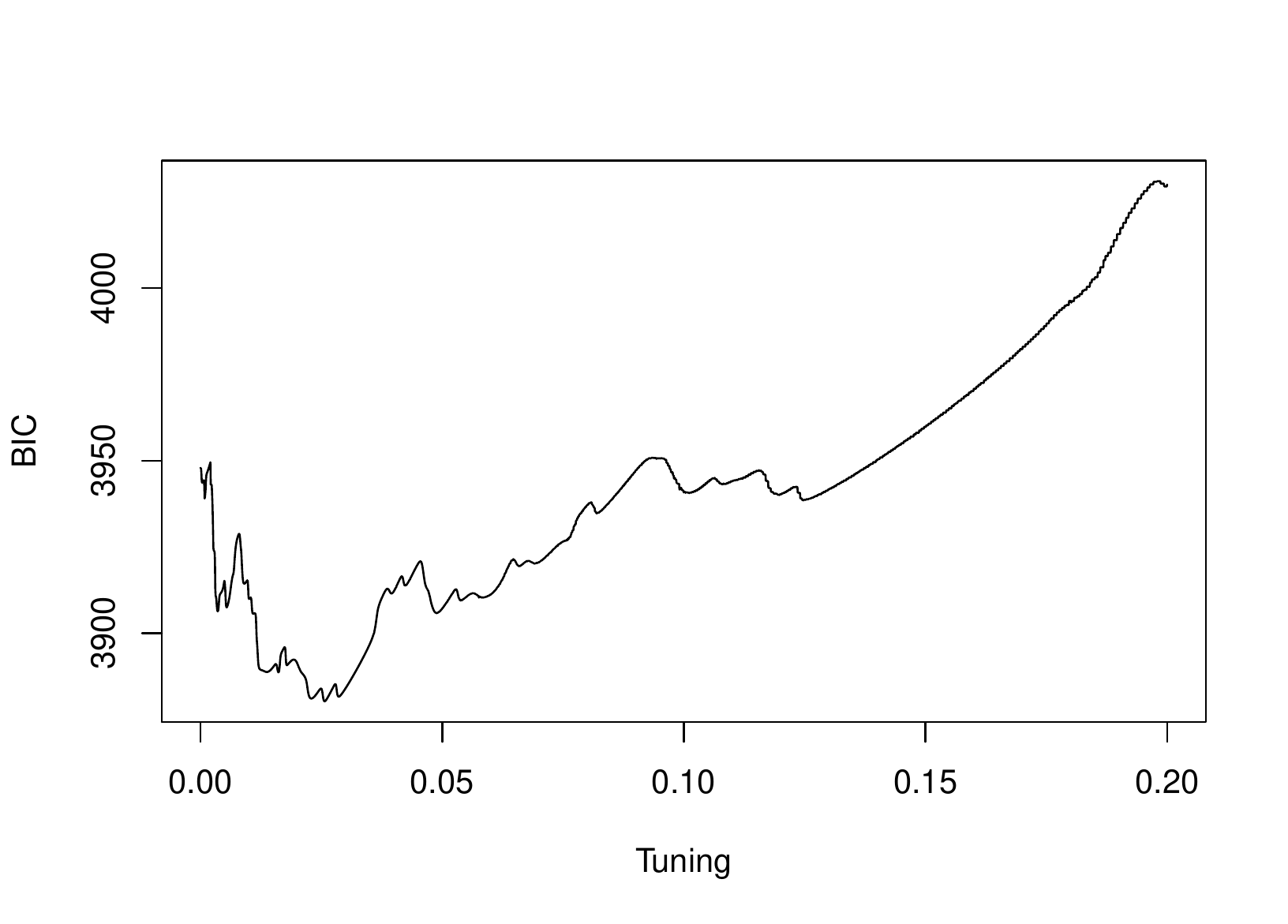}}}\\
             %\label{fig:BIC_1}}\\
             \subfloat[]{{\includegraphics[scale=0.6]{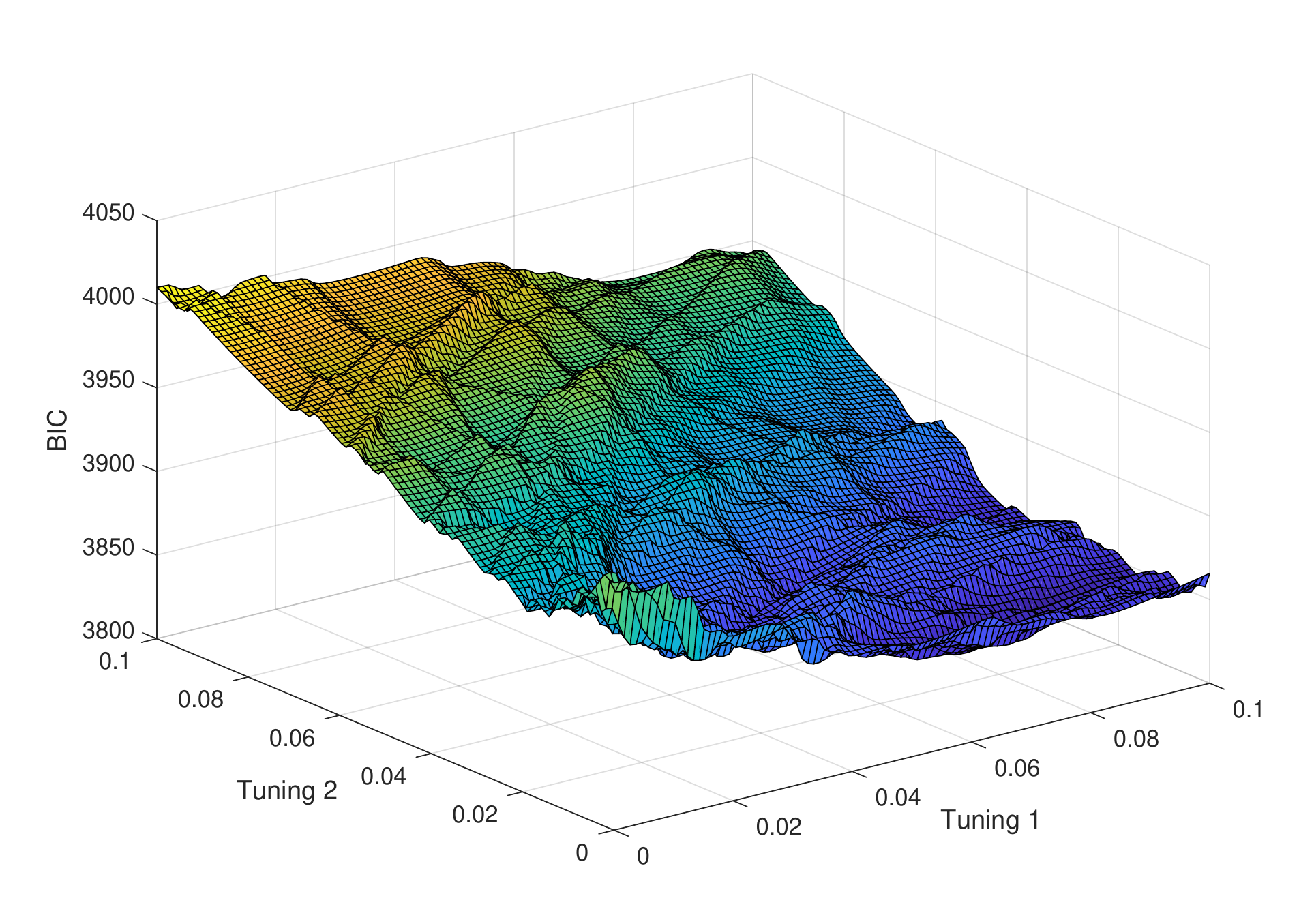}}} %\label{fig:BIC_2}}
             \caption{The BIC function evaluated at different tuning parameter values for the Weibull MPR model with the LASSO penalty for the lung cancer dataset analysed in Section \ref{Sec:Data}, (a) corresponds to one tuning parameter and (b) corresponds to two.\label{fig:BIC}}
			\end{center}
	 \end{figure}

		The simplest method to solve this optimization problem is grid search. While it is straightforward to implement, grid search is known to suffer from the curse of dimensionality, i.e., the number of grid points grows exponentially with the dimension. %We've found that the algorithm becomes inefficient even in the 2-d space.
		Furthermore, if the grid is too coarse, the minimum may be overlooked. This is especially true in the case of a multi-modal function, such as what we are trying to optimize (see Figure \ref{fig:BIC}). To overcome the issues associated with the grid search algorithm, we consider ``global'' optimization algorithms. In an empirical comparison of various algorithms for continuous global optimization, \citet{mullen2014continuous} found ``\texttt{DEoptim}'' (implemented in \texttt{R}) \citep{mullen2009deoptim} to be among the best. The function implements a differential evolution algorithm, an example of an evolutionary strategy developed by \citet{storn1997differential} (see \citet{mullen2009deoptim} for a detailed overview of the underlying algorithm). %Differential evolution algorithms are inherently parallel, hence reducing computation time considerably.
		%These algorithms are particularly well-suited to find the global optimum of a non-monotic, real-valued function with real-valued parameters. A further significant speedup can be obtained if the algorithm is executed on a parallel machine or a network ofcomputers.
	
{\subsection{Variable Selection Algorithm}
The variable selection algorithm described above is summarized in the following bullet points.
\begin{itemize}
\item {\it Initialization.} Set $\theta^{(0)} = \hat\theta_0$ where $\hat\theta_0$ is the vector of unpenalized estimates, i.e., those which minimize $\ell(\theta)$ (defined in (\ref{eq:NonPen_lik})).
\item {\it  Optimization.}
\begin{itemize}
\item {\it  Outer.} Minimize $\text{BIC}(\lambda)$ with respect to $\lambda$ using \texttt{DEoptim}, yielding $\lambda^*$ as defined in (\ref{eq:lambdastr}). Convergence occurs when $|\text{BIC}(\lambda^{(r+1)})-\text{BIC}(\lambda^{(r)})|$ is below a prespecified threshold. (Here, $\lambda^{(r)}$ is the best $\lambda$ value found at step $r$ of the \texttt{DEoptim} algorithm.)
\item {\it Inner.} For a given value of $\lambda$, maximize $\ell_\lambda(\theta)$ by iteratively re-solving the system of equations given in (\ref{eq:NR}) starting from the initial value, $\theta^{(0)}$; this yields $\hat\theta_\lambda$. Convergence occurs when $||\theta_\lambda^{(m+1)}-\theta_\lambda^{(m)}||_\infty$ is below a prespecified threshold. (Here $||y||_\infty = \max_j | y_j |$ is the infinity norm.)
\end{itemize}
\item {\it Output.} The estimates $\hat\theta_{\lambda^*}$ are returned from the above procedure, and the corresponding standard errors are calculated by evaluating (\ref{eq:SE_formula}) at $\hat\theta_{\lambda^*}$.
\end{itemize}
}

\section{Simulation Studies}
\label{sec:Simulations}
\subsection{Setup}%Data Generation}
	The performance of the proposed variable selection methods is evaluated through simulation studies. The failure time is simulated from a Weibull MPR model with
	$$\log(\tau_i) = x^{T}_{i}(-1.5, -1.0, 0.0, 0.0, 0.0, 0.0, 0.0, -0.8, 0.5, 0.0, 0.0)^{T},$$
	$$\log(\gamma_i) = z^{T}_{i}(0.5, 0.4, 0.0, 0.0, 0.0, 0.4, -0.2, 0.0, 0.0, 0.0, 0.0)^{T},$$
	where $x_i = z_i = (1,x_{i1},\dots,x_{i10})^T$ is a vector of correlated variables generated from an AR(1) process with a correlation coefficient $\rho = 0.5$. Each variable is marginally standard normal and the correlation between any two consecutive variables $x_{ij}$ and $x_{ik}$ is given by $\rho^{|j-k|}$. The corresponding censored times were generated from an exponential distribution. This setup was chosen so as to yield realistic survival data, where the true model is sparse and correlation in the covariates exists. Three different sample sizes ($n$ = 100, 500 and 1000) and two censoring proportions ($p_{\text{cen}} = 25\%, 50\%$) are considered. For each scenario, we considered the LASSO, SCAD, and ALASSO penalties with both a single tuning parameter or two tuning parameters (i.e., one for each of the two regression components). Each simulation was repeated 200 times.
	\subsection{Simulation Results}
    The variable selection and estimation procedures described in Sections \ref{sec:Formulation} and \ref{sec:Estimation} are applied to the simulated data and the results are summarized and discussed here. A number of metrics are used to evaluate the performance of the variable selection procedures, namely: the average number of true zero coefficients \emph{correctly} set to zero (C), the average number of true non-zero coefficients \emph{incorrectly} set to zero (IC), and the probability of choosing the true model (PT); for the oracle model, C = 7 and IC = 0. As a measure of prediction accuracy, we also consider the mean squared error (MSE), given by {$\text{MSE}(\hat{\beta}) = (\hat{\beta} - \beta)^{T} V (\hat{\beta} - \beta)$ and $\text{MSE}(\hat{\alpha}) = (\hat{\alpha} - \alpha)^{T} V (\hat{\alpha} - \alpha)$, where $V$, the simulated sample covariance matrix of the covariates, is computed for each simulation replicate \citep{zhang2007adaptive, tibshirani1997lasso}}. These metrics, averaged over simulation replicates for the scenarios with 25\% censoring, are reported in Table \ref{tab:SimResults}. (The results for 50\% censoring, shown in the Appendix, are similar.)
	\par
	As the sample size increases, we see an improvement across all the four metrics, for both the shape and the scale parameters and across all penalties. However, it is evident that the LASSO penalty does not set enough covariates equal to zero (i.e., it selects an overly complex model) irrespective of whether there is one tuning parameter or two. SCAD performs better but over-selects in the shape component, $\alpha$, when there is only one tuning parameter; this is improved by having two separate tuning parameters. The best performance comes from the ALASSO penalty which, for the largest sample size, selects the true scale and shape covariates more than 90\% of the time. Interestingly, the ALASSO performs well even with a single tuning parameter (but it does improve a little with two tuning parameters). Computation times for each of the penalties are given in Table \ref{tab:Comptime} where we see that SCAD is considerably slower than LASSO and ALASSO, while the times for these latter two are comparable. Furthermore, the computation times for the cases with two tuning parameters are 2 - 3 times longer than those with one tuning parameter.
	\par 	
Besides variable selection, we also consider parameter inference in terms of estimation bias, accuracy of the estimated standard error (SEE) computed using the sandwich formula, (\ref{eq:SE_formula}), and { the empirical coverage probability (CP) of a nominal 95\% confidence interval}. The results for the ALASSO penalty (for the 25\% censoring level) are presented in Table \ref{tab:MSEcheck}. Results for the ALASSO penalty show that the SEE is accurate for moderate sample sizes, but may underestimate the standard error (SE) for smaller samples. In the samples $n = 100$ and $n = 500$, the smaller parameters are overshrunk, i.e., they are biased downwards. For this reason the CPs do not perform well. However, this is not the case for $n = 1000$. In the case of $n = 1000$, {the CPs are close to the nominal $95\%$ level} (although perhaps a little low for the shape coefficients). An improvement can be seen in the case of two tuning parameters. Similar tables for the other penalties (and 50\% censoring) can be found in the Appendix, and, to summarize these additional results: LASSO overshrinks all parameters and the standard errors are underestimated in all cases, whereas the results for SCAD are better, but not as good as the ALASSO; in particular, the coverage for SCAD confidence intervals is much lower than the nominal level even for $n=1000$. { As expected, when the censoring rate is increased from 25\% to 50\%, the variation (i.e., SE, SEE and MSE) of estimates is increased overall across all the three penalties.}
	\begin{table}%[!ht]
     \caption{Simulation results: variable selection metrics averaged over 200 simulation replicates.\label{tab:SimResults}}	
	 \begin{center}	
     \begin{adjustbox}{width=1\textwidth}
		 \begin{threeparttable}
      {\footnotesize \begin{tabular}{cr|cccc|cccc|cccc}
					\hline
                 \multicolumn{2}{c}{}&\multicolumn{12}{c}{One Tuning Parameter}\\
					\hline
					{} & {} & \multicolumn{4}{c|}{LASSO} &\multicolumn{4}{c|}{SCAD} &\multicolumn{4}{c}{ALASSO}\\
					$p_{\text{cen}}$ = 25\% & $n$ & C(0) & IC(7)& PT & MSE &C(7) & IC(0)&PT& MSE &C(7) & IC(0)&PT & MSE\\
					\hline
					{}&100& 5.29 & 0.18 & 0.12 & 0.33
					&6.32 &  0.14 &  0.53 & 0.33
					& 6.17 & 0.05 & 0.38 & 0.19  \\
					Scale ($\beta$)&500 &5.52 & 0.00 & 0.18 & 0.09
					&6.96 & 0.00 & 0.98 & 0.03
					& 6.79 & 0.00 & 0.83 & 0.03\\
					{} & 1000 & 5.83 & 0.00 & 0.31 & 0.05
					&6.99 & 0.00  & 0.99 & 0.01
					& 6.93 & 0.00 &0.94 & 0.01 \\
					
					\hline
					{}&100& 3.40 &  0.06 & 0.00 & 0.05
					& 4.19 & 0.09 & 0.02 & 0.05
					& 6.25 & 0.16 & 0.40 &  0.04 \\
					Shape ($\alpha$)& 500 & 3.37 &  0.00 & 0.01 & 0.01
					& 5.86  &0.00 & 0.33 & 0.01
					&  6.81 & 0.00 & 0.85 & 0.00 \\
					{} & 1000 & 3.46 & 0.00 & 0.00 & 0.01
					&6.51 & 0.00 & 0.64 & 0.00
					& 6.93 & 0.00 & 0.95 & 0.00 \\
					\hline
					\multicolumn{2}{c}{}&\multicolumn{12}{c}{Two Tuning Parameters}\\

					\hline
					{} & {} & \multicolumn{4}{c|}{LASSO} &\multicolumn{4}{c|}{SCAD}&\multicolumn{4}{c}{ALASSO}\\
					$p_{\text{cen}}$ = 25\% & $n$ & C(7) & IC(0) & PT & MSE &C(7) & IC(0)&PT& MSE &C(7) & IC(0)&PT & MSE\\
					\hline
					{}&100& 4.61 & 0.10 & 0.06 & 0.28
					& 6.26 & 0.10 & 0.55 & 0.28
					& 6.47 & 0.14 & 0.52 & 0.23 \\
					Scale ($\beta$)&500 & 5.40 & 0.00 & 0.20 & 0.08
					& 6.95 & 0.00& 0.97 & 0.02
					& 6.84 & 0.00 &0.88 & 0.03 \\
					{} & 1000 & 5.32 & 0.00 & 0.20 & 0.04 & 6.89& 0.00& 0.96& 0.01 &6.92 & 0.00& 0.92 & 0.01\\
					\hline
					{}&100& 5.27 & 0.25 &0.10 & 0.05
					& 5.15 & 0.20 & 0.11 & 0.06
					&6.30 & 0.25 & 0.44 & 0.05\\
					Shape ($\alpha$)&500 & 5.30 & 0.00 & 0.17 & 0.01
					& 5.98 & 0.00 & 0.36 & 0.01
					&6.93 & 0.00 & 0.93 & 0.00\\
					{} & 1000 & 5.50 & 0.00& 0.26 & 0.01 & 6.51& 0.00 &0.63 & 0.00 & 6.94 &0.00 & 0.94 & 0.00\\
					\hline
\end{tabular}}
{\footnotesize\begin{tablenotes}
\item[]{ {C, average correct zeros; IC, average incorrect zeros; PT, the probability of choosing the true model; MSE, the average mean squared error.}}
\end{tablenotes}}
\end{threeparttable}
\end{adjustbox}
 \end{center}
\end{table}

	\begin{table}%[!ht] \centering \footnotesize
     \caption{Average computation time per simulation replicate (in minutes).\label{tab:Comptime}}
     \begin{center}
				{\footnotesize \begin{tabular}{cr|ccc|ccc}
					\hline
					\multicolumn{1}{c}{} & \multicolumn{1}{c}{} &\multicolumn{3}{c}{One Tuning Parameter} &\multicolumn{3}{c}{Two Tuning Parameters}\\
					\hline
					$p_{\text{cen}}$ = 25\% & $n$&LASSO&SCAD&ALASSO&LASSO&SCAD&ALASSO\\
					\hline
					{}&100& 1.5 & 6.1 & 1.9 & 4.6& 11.7 & 4.7\\
					{}&500 & 2.7 & 6.2 & 3.2 & 7.3& 19.9 & 9.0\\
					& 1000 & 4.3 & 11.5 & 5.8 & 13.1 & 37.5 & 17.1 \\
					\hline
     \end{tabular}}
   \end{center}
	\end{table}

	\begin{table}%[!ht] \centering \footnotesize
     \caption{ Further simulation results: estimates, standard errors, and confidence intervals.\label{tab:MSEcheck}}
	 \begin{center}
		\begin{adjustbox}{width=1\textwidth}
		\begin{threeparttable}
				{\footnotesize \begin{tabular}{cc|cccc|cccc|cccc}
					\hline
					\multicolumn{14}{c}{ALASSO}\\
					\hline
					\multicolumn{14}{c}{One Tuning Parameter}\\
					\hline
					$p_{\text{cen}}$ = 25\% & & \multicolumn{4}{c|}{$n$ = 100} & \multicolumn{4}{c|}{$n$ = 500} & \multicolumn{4}{c}{$n$ = 1000}\\
					{} & $\theta$ & $\hat{\theta}$ & SE & SEE & CP & $\hat{\theta}$ & SE & SEE & CP & $\hat{\theta}$ & SE & SEE & CP\\
					\hline
					$\beta_0$ & -1.50 & -1.49 & 0.21 & 0.21 & 0.96
					& -1.47 & 0.09 & 0.09 & 0.92
					& -1.48 & 0.06 & 0.06 & 0.95\\
					$\beta_1$ & -1.00 & -0.99 & 0.19 & 0.17 & 0.93
					& -0.98 & 0.07 & 0.07 & 0.93
					&  -0.99 & 0.05 & 0.05 & 0.95\\
					$\beta_7$ & -0.80& -0.76 & 0.19 & 0.15 & 0.87
					& -0.77 & 0.07 & 0.06 & 0.89
					& -0.79 & 0.05 & 0.04 & 0.94\\
					$\beta_8$ & 0.50 & 0.42 & 0.18 & 0.13 & 0.85
					& 0.47 & 0.06 & 0.06 & 0.88
					& 0.48 & 0.04 & 0.04& 0.91 \\
					\hline
					$\alpha_0$ & 0.50 & 0.54 & 0.10 & 0.09&  0.92
					&0.50 &  0.04& 0.04 & 0.95
					& 0.50& 0.03& 0.03& 0.98\\
					$\alpha_1$ & 0.40 & 0.38 & 0.06& 0.06& 0.93
					& 0.40 & 0.02 & 0.02 & 0.93
					& 0.40& 0.02& 0.01& 0.91\\
					$\alpha_5$ & 0.40 & 0.35 & 0.08 & 0.06& 0.77
					& 0.39& 0.03 & 0.02& 0.91
					& 0.39& 0.02& 0.02& 0.91\\
					$\alpha_6$ &-0.20 & -0.15 &0.09 &0.05&0.77
					& -0.18 & 0.03 & 0.02& 0.87
					& -0.19& 0.02& 0.02& 0.91 \\
					\hline
					\multicolumn{14}{c}{Two Tuning Parameters}\\
					\hline
					$p_{\text{cen}}$ = 25\% & & \multicolumn{4}{c|}{$n$ = 100} & \multicolumn{4}{c|}{$n$ = 500} & \multicolumn{4}{c}{$n$ = 1000}\\
					{} & $\theta$ & $\hat{\theta}$ & SE & SEE & CP & $\hat{\theta}$ & SE & SEE & CP & $\hat{\theta}$ & SE & SEE & CP\\
					\hline
					$\beta_0$ & -1.50 & -1.48 & 0.24 & 0.21 & 0.92
					& -1.49 & 0.10 & 0.09 & 0.93
					& -1.50 & 0.06 & 0.06 & 0.97\\
					$\beta_1$ & -1.00 & -0.97 & 0.22 & 0.16 & 0.86
					& -0.99 & 0.08 & 0.07 & 0.92
					&  -1.00 & 0.05 & 0.05 & 0.95\\
					$\beta_7$ & -0.80&  -0.74 & 0.21 & 0.15 & 0.86
					& -0.79 & 0.07 & 0.06 & 0.93
					& -0.79 & 0.05 & 0.04 & 0.92\\
					$\beta_8$ & 0.50 & 0.38 & 0.20 & 0.12 & 0.77
					& 0.47 & 0.07 & 0.06 & 0.87
					& 0.49 & 0.04 & 0.04 & 0.93 \\
					\hline
					$\alpha_0$ & 0.50 & 0.53 & 0.11 & 0.09&  0.89
					&0.50 &  0.05& 0.04 & 0.92
					& 0.50& 0.03& 0.03& 0.97\\
					$\alpha_1$ & 0.40 & 0.37 & 0.07& 0.06& 0.86
					& 0.39 & 0.02 & 0.02 & 0.95
					& 0.40& 0.01& 0.01& 0.95\\
					$\alpha_5$ & 0.40 & 0.35 & 0.10 & 0.06& 0.72
					& 0.39 & 0.03 & 0.02& 0.93
					& 0.40& 0.02& 0.02& 0.95\\
					$\alpha_6$ &-0.20 & -0.14 &0.10 &0.05&0.72
					& -0.19 & 0.03 & 0.02& 0.87
					& -0.19& 0.02& 0.02& 0.93 \\
					\hline
     \end{tabular}}
   {\footnotesize\begin{tablenotes}
	\item[]{ SE, standard deviation of estimates over 200 replications; SEE, average of estimated standard errors over 200 replications; CP, the empirical coverage probability of a nominal 95\% confidence interval.}
	\end{tablenotes}}
\end{threeparttable}
\end{adjustbox}
 \end{center}
\end{table}

 \section{Lung Cancer Data}
 \label{Sec:Data}
 To illustrate the penalized variable selection methods on real data, a lung cancer dataset is considered. This dataset was collected as part of a PhD thesis by \citet{wilkinson1995lungc} (see also \citet{burke2017multi}). This dataset contains all individuals, of all ages, diagnosed with lung cancer in Northern Ireland during the one year period 1 October 1991 to 30 September 1992. Only cases of primary lung cancer were included. The date of diagnosis was taken to be the time origin for an individual and the end point was the earlier of the occurrence of death or the study end date, which was on 30 May 1993. Individuals who were still alive on the study end date were taken to have censored survival times. Individuals who died from another cause or who dropped out of the study were also censored. The final dataset included 855 patients, of which there were 673 deaths and 182 censored times. Besides the survival time and the censoring indicator, a number of other variables were recorded for each of the patients enrolled in the study (reference categories are listed first): age group ($< 40-$, $50-$, $60-$, $70-$, $> 80$), sex (female, male), treatment group (palliative, surgery, chemotherapy, radiotherapy, chemotherapy and radiotherapy), WHO status (normal activity, light work, unable to work, $> 50\%$ walking, bed/chair bound), cancer cell type (squamous cell, small cell, adenocarcinoma, other), serum sodium level ($\ge 136\ mmol/l$, $< 136\ mmol/l$, missing), serum albumen level ($\ge 35\ g/l$, $< 35\ g/l$, missing), metastases (no, yes, unknown), smoking status (non-smoker, current smoker, ex-smoker, missing).

 \subsection{Adequacy of Weibull}

{
Before considering covariates and variable selection, we first carry out an initial check that a baseline Weibull distribution is appropriate for the lung cancer data. The cumulative hazard function for the Weibull model is given by $H(t) = \int_0^t h(u) du = \lambda t^\gamma$, and, hence, $\log H(t) = \log \lambda + \gamma \log t$. Therefore, given an estimate $\hat H(t)$, a plot of $\log \hat H(t)$ against $\log t$ should produce a straight line. This standard Weibull model check is shown in Figure \ref{fig:weibullfit}, and, despite a slight deficiency for very small survival times, it appears that the Weibull model is reasonable.
\begin{figure}[!ht]
   \begin{center}
     \includegraphics[scale=0.5]{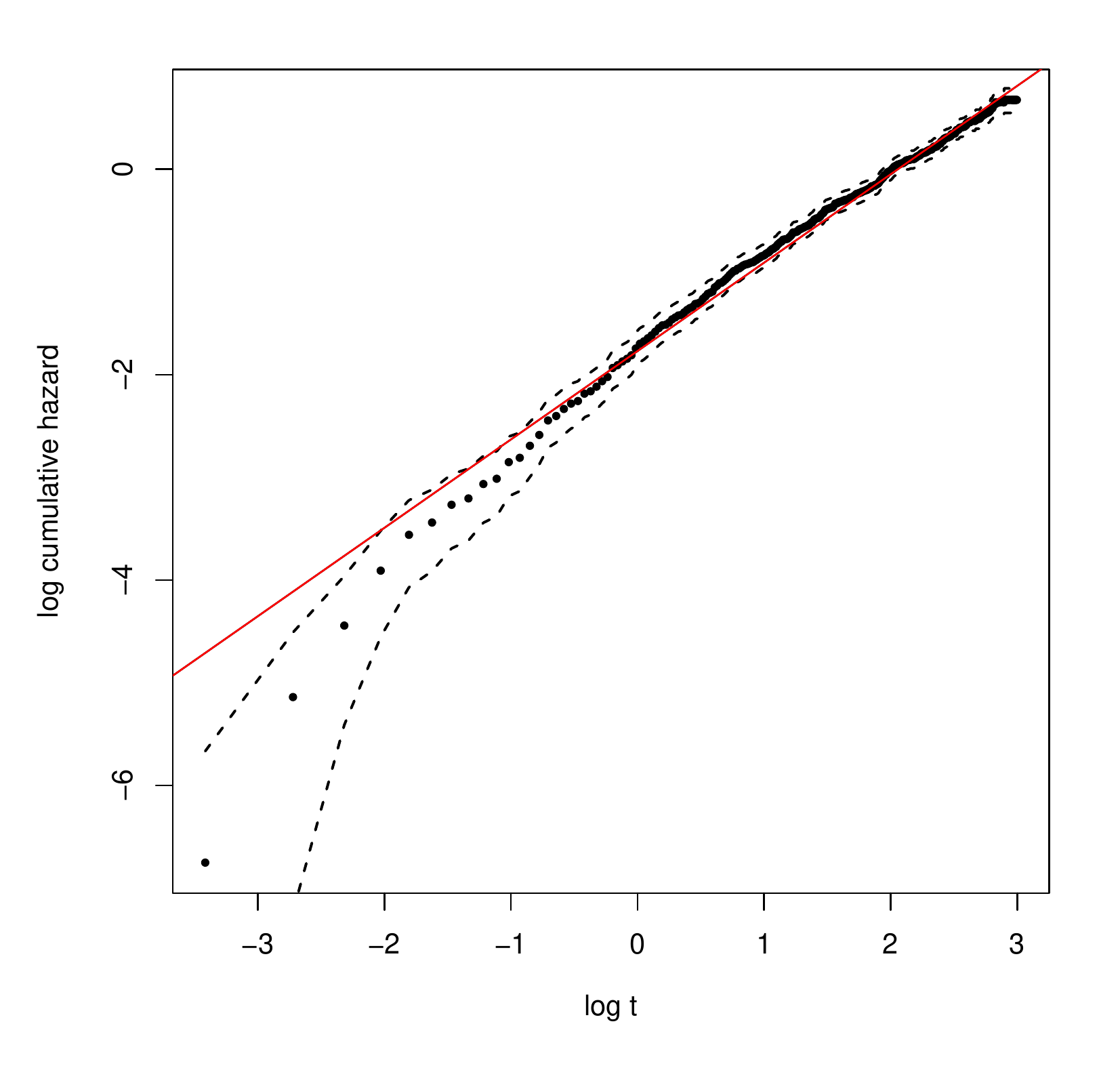}
     \caption{Weibull model check. Here $\hat H(t)$, along with the 95\% confidence intervals, come from the Kaplan-Meier estimator.\label{fig:weibullfit}}
   \end{center}
\end{figure}
}

 \subsection{Variable Selection Results}
     The variable selection results for the different penalties are summarized in Table \ref{tab:LC_VSsummary}. In line with the results of the simulation study, the LASSO penalty selects the most complex model and the ALASSO penalty selects the least complex. Both ALASSO penalties (one and two tuning parameter cases) are in agreement on the non-importance of sex and smoking status, and although age group is selected in the scale in the case with one tuning parameter, it is not significant. {Interestingly, the two tuning parameter ALASSO selects the same set of covariates as identified by \citet{burke2017multi} using a BIC stepwise procedure (albeit they additionally selected treatment in the shape)}. We also see that, in the two tuning parameters cases, the scale tuning parameter is smaller that of the one tuning parameter case, while the shape tuning parameter is larger. This suggests that the single penalty over-penalizes the scale coefficients and under-penalizes the shape; this is also evident from the scale and shape degrees of freedom. Interestingly, the one tuning parameter ALASSO converges in less than half the time of the two tuning parameter ALASSO, and achieves similar results. We expect this based on our simulation studies, and also expect the results of the two tuning parameter case to be marginally better (albeit it takes longer to converge).
     \par
	Table \ref{tab:LC_VSestA-lasso} displays the estimated coefficients for both ALASSO penalties along with the unpenalized coefficients (we focus the ALASSO due to its superior performance in our simulation studies, but similar tables for LASSO and SCAD can be found in the Appendix). From \citet{burke2017multi}, note that the scale coefficients characterize the overall scale of the hazard (a positive value indicates an increase relative to the reference category), while the shape coefficients characterize its time evolution (a positive value indicates a hazard which increases over time relative to the reference category). We clearly see the similarity of the coefficient values for both the one and two tuning parameter ALASSO penalties, and, furthermore, that the selected variables are broadly in line with those which are statistically significant in the unpenalized model. Focussing on the results of the two tuning parameter case we find that: all treatments (apart from chemotherapy) have a negative scale coefficient suggesting that treatment reduces hazard (relative to palliative care); however, worse WHO status, small cancer cell type, presence of metasteses, and reduced sodium and albumen levels increase the hazard; lastly, sex, age group, and smoking status have no significant effect on the hazard. Since no variable appears in the shape component (i.e., all shape coefficients are set to zero), the selected model is a proportional hazards model, and exponentiating the scale coefficients yields the hazard ratios, e.g., the surgery hazard ratio is $\exp(-0.98) = 0.375$ so that the risk of death is approximately 37.5\% that of a patient receiving palliative care.
     \begin{table}[!ht]
         \caption{Summary of penalized models {(lung cancer dataset)} \label{tab:LC_VSsummary}}
         \begin{center}
         \begin{threeparttable}
         {\footnotesize\begin{tabular}{l|ccc|ccc}
           \hline
           {} & \multicolumn{3}{c|}{One Tuning Parameter}& \multicolumn{3}{c}{Two Tuning Parameters} \\
           {} & LASSO & SCAD & ALASSO & LASSO & SCAD & ALASSO\\
           \hline
           Treatment & $\boldsymbol{\beta}$, {\color{gray} $\alpha$} & \bm{$\beta$}, \bm{$\alpha$} & \bm{$\beta$}, \bm{$\alpha$} &\bm{$\beta$}, {\color{gray}$\alpha$} &\bm{$\beta$}, {\color{gray}$\alpha$} & \bm{$\beta$} \\
           Age group & {\color{gray} $\alpha$}  & {\color{gray}$\alpha$} & {\color{gray} $\beta$} & {\color{gray} $\alpha$} & {\color{gray} $\alpha$} & - \\
           WHO status &\bm{$\beta$}, {\color{gray}$\alpha$} & \bm{$\beta$}, \bm{$\alpha$} & \bm{$\beta$} & \bm{$\beta$}, {\color{gray} $\alpha$} & \bm{$\beta$}, {\color{gray}$\alpha$} & \bm{$\beta$} \\
           Sex & {\color{gray} $\alpha$} & - & - & - & - & -  \\
           Smoking status & {\color{gray} $\alpha$} & {\color{gray}$\alpha$} & - & \bm{$\beta$} & \bm{$\beta$} & - \\
           Cell type  & \bm{$\beta$}, {\color{gray} $\alpha$} & \bm{$\beta$}, {\color{gray} $\alpha$} & \bm{$\beta$} & \bm{$\beta$} & \bm{$\beta$}, {\color{gray} $\alpha$} & \bm{$\beta$} \\
           Metastases & \bm{$\beta$}, {\color{gray} $\alpha$} & \bm{$\beta$}, {\color{gray} $\alpha$} & \bm{$\beta$} & \bm{$\beta$} & \bm{$\beta$} & \bm{$\beta$}\\
           Sodium & \bm{$\beta$}, {\color{gray} $\alpha$} & \bm{$\beta$}, {\color{gray} $\alpha$}  & {\color{gray} $\beta$} & \bm{$\beta$} & \bm{$\beta$} & \bm{$\beta$}\\
           Albumen & \bm{$\beta$}, \bm{$\alpha$} &\bm{$\beta$}, {\color{gray} $\alpha$} & \bm{$\beta$} & \bm{$\beta$}, {\color{gray} $\alpha$} & \bm{$\beta$}, {\color{gray} $\alpha$} & \bm{$\beta$} \\[15pt]
         %%%%%%%%%%%%%%%%%%%%%%%%%%%%%%%%%%%%%%%%%%%%%%%%%%%%%%%%%%%%%%
           Tuning parameter(s) & 0.026 & 0.041 & 0.015 & 0.014 & 0.024 & 0.004 \\
					& & & & 0.080 &  0.074 & 0.045 \\
					Degrees of freedom & 32.5 & 27.1 & 15.5 &
				 25.6 & 24.6 & 15.2\\
					Scale degrees of freedom & 14.2 & 12.0 & 13.2 &
				 18.3 & 17.1 & 14.2 \\
					Shape degrees of freedom & 18.3 & 15.0 & 2.4 &
				 7.4 & 7.6 & 1.0\\
           Computation time (in minutes)& 27.0 & 41.9 & 28.8 & 58.3 & 100.6 & 86.9 \\
           \hline
         \end{tabular}}
           {\footnotesize\begin{tablenotes}
           \item[] $\beta$ = ``selected in scale'', $\alpha$ = ``selected in shape'', and those which are non-significant (at the $5\%$ level) are shown in gray.
           \end{tablenotes}}
           \end{threeparttable}
       \end{center}
	 \end{table}
	
		\begin{sidewaystable}%[!ht]
         \caption{Coefficients estimates and standard errors for the ALASSO penalties {(lung cancer dataset)} \label{tab:LC_VSestA-lasso}}\centering
             \begin{threeparttable}
					\begin{adjustbox}{max width=\linewidth}
					{\footnotesize	\tabcolsep = 12.9pt	\begin{tabular}{ll|ccc|ccc}
						\hline
						{} & {} & \multicolumn{3}{c|}{Scale} & \multicolumn{3}{c}{Shape}\\
						\hline
						\multicolumn{2}{c|}{Covariate} & No Penalty & One Tuning &Two Tuning& No Penalty& One Tuning & Two Tuning\\
						Intercept & & \bf{-3.38 (0.66)} & \bf{-3.12 (0.17)} & \bf{-3.15 (0.17)} &  -0.16 (0.22) & 0.04 (0.03) & 0.04 (0.03)\\
						Treatment & surgery & \bf{-1.69 (0.83)}& \bf{-0.89 (0.21)} & \bf{-0.98 (0.22)} & 0.11 (0.21) & 0.00 (0.00) & 0.00 (0.00) \\
						& chemotherapy &  -0.33 (0.37) & 0.00  (0.00)  & 0.00 (0.00) & -0.03 (0.15) & 0.00 (0.00) & 0.00 (0.00) \\
						& radiotherapy & \bf{-0.85 (0.21)} & -0.16 (0.10) & \bf{-0.21 (0.10)} & \bf{0.22 (0.08)} & 0.00 (0.00) &  0.00 (0.00) \\
						& chemo. \& radio. & \bf{-3.83 (0.98)}& \bf{-2.30 (0.89)} & \bf{-0.63 (0.22)} & \bf{0.77 (0.20)} & \bf{0.51 (0.21)} & 0.00 (0.00)\\
						Age group & $50-$ &  \bf{-0.90 (0.43)} & 0.00 (0.00) & 0.00 (0.00) & \bf{0.39 (0.16)} & 0.00 (0.00) & 0.00 (0.00)\\
						& $60-$ & \bf{-0.94 (0.39)} &0.00 (0.00) & 0.00 (0.00) & \bf{0.40 (0.15)} & 0.00 (0.00) & 0.00 (0.00)\\
						& $70-$ & -0.77 (0.39) &  0.02 (0.08) & 0.00 (0.00)& \bf{0.31 (0.15)} & 0.00 (0.00) & 0.00 (0.00)\\
						& $> 80$ & -0.78 (0.42) & 0.00 (0.00)  & 0.00 (0.00) & 0.31 (0.17) & 0.00 (0.00) & 0.00 (0.00)\\
						WHO status & light work & -0.02 (0.45) & 0.00 (0.00) & 0.00 (0.00) & 0.02 (0.12) & 0.00 (0.00) & 0.00 (0.00) \\
						& unable to work & 0.84 (0.43) & \bf{0.41 (0.10)} & \bf{0.44 (0.10)} &-0.10 (0.13) & 0.00 (0.00) & 0.00 (0.00) \\ 	
						& $> 50\%$ walking & \bf{1.31 (0.44)} & \bf{0.99 (0.11)} & \bf{0.97 (0.11)} & -0.13 (0.14) & 0.00 (0.00) & 0.00 (0.00)\\
						& bed/chair bound & \bf{1.78 (0.50)} & \bf{1.28 (0.28)} & \bf{1.54 (0.25)}&  -0.03 (0.20) & 0.00 (0.00) & 0.00 (0.00) \\
						Sex & male & 0.03 (0.14) & 0.00 (0.00) &0.00 (0.00) & -0.03 (0.05) & 0.00 (0.00) & 0.00 (0.00) \\
						Smoking status & current smoker & 0.10 (0.22) & 0.00 (0.00) & 0.00 (0.00) & 0.15 (0.08) & 0.00 (0.00) & 0.00 (0.00) \\
						& ex-smoker & -0.05 (0.23) & 0.00 (0.00) & 0.00 (0.00) & 0.17 (0.09) & 0.00 (0.00) & 0.00 (0.00) \\
						& missing &  0.29 (0.40) & 0.00 (0.00) & 0.00 (0.00) &0.00 (0.00)  & 0.00 (0.00) & 0.00 (0.00) \\
						Cell type & small cell &  \bf{0.83 (0.26)} &  \bf{0.31 (0.12)} & \bf{0.43 (0.13)} & -0.05 (0.10) & 0.00 (0.00) & 0.00 (0.00) \\
						& adenocarcinoma & 0.28 (0.28) & 0.00 (0.00) & 0.00 (0.00) & 0.03 (0.10) & 0.00 (0.00) & 0.00 (0.00) \\
						& other & 0.32 (0.20) & 0.00 (0.00) & 0.09 (0.09) & -0.04 (0.07) & 0.00 (0.00) & 0.00 (0.00)  \\
						Metastases & yes &  \bf{1.35 (0.28)} & \bf{0.89 (0.12)} & \bf{0.84 (0.12)}& \bf{-0.19 (0.08)} & 0.00 (0.00) & 0.00 (0.00) \\
						& unknown &   \bf{0.83 (0.30)} & \bf{0.53 (0.13)} & \bf{0.41 (0.13)} & -0.14 (0.09) & 0.00 (0.00) & 0.00 (0.00)\\
						Sodium level & $< 136\ mmol/l$ &  \bf{0.33 (0.14)} & 0.14 (0.08) & \bf{0.24 (0.08)}& -0.01 (0.05) & 0.00 (0.00) & 0.00 (0.00)\\
						& missing &  -0.77 (0.45) & 0.00 (0.00) & 0.00 (0.00)&  \bf{0.32 (0.16)} & 0.00 (0.00) & 0.00 (0.00) \\
						Albumen level & $< 35\ g/l$ &  \bf{0.65 (0.16)} & \bf{0.36 (0.09)} &  \bf{0.37 (0.09)} & -0.10 (0.06) & 0.00 (0.00) & 0.00 (0.00) \\
						& missing & \bf{0.59 (0.28)} & 0.00 (0.00) &  \bf{0.27 (0.14)}& 0.09 (0.15) & 0.00 (0.00) & 0.00 (0.00) \\
						\hline
					\end{tabular}}
       \end{adjustbox}
       {\footnotesize\begin{tablenotes}
         \item[] Bold indicates statistically significant at the $5\%$ level.
       \end{tablenotes}}
       \end{threeparttable}
\end{sidewaystable}
\section{Discussion}
\label{sec:conc}
The MPR approach results in flexible models which extend standard models, but the presence of multiple regression components means that variable selection is necessarily more challenging than in standard settings where there is only a single regression component. In this article, we have proposed a penalized variable selection procedure for the simultaneous selection of significant variables in the shape and scale parameters of a Weibull MPR model in the survival analysis setting. The performance of these methods was illustrated using simulation studies and a real data example. While we have considered the Weibull model example in this article, the proposed variable selection procedures can be applied easily to other MPR models.
\par
Given that we model different distributional parameters (a scale and a shape parameter), there is no reason to assume that variable selection can be achieved with a single penalty applied to both regression components; hence, we also investigated the need for a separate tuning parameter for each regression component. We have found that the ALASSO performs very favourably in terms of identifying the true subset of covariates and coverage of calculated confidence intervals. This is true even with a single tuning parameter, however the results are improved when there are two tuning parameters (albeit this is more computationally intensive). On the other hand, SCAD does not perform well in the MPR setting, selecting an overly complex model and with poor confidence interval coverage for shape parameters.

\section*{Acknowledgements}

The first author is funded by the the Irish Research Council. The second author was supported by the Basic Science Research Program
through the National Research Foundation of Korea (NRF) funded by the Ministry of Science \& ICT (No.~NRF-2017R1E1A1A03070747)

\bibliographystyle{apa}
\bibliography{Bibliography-MM-MC}

\newpage

\section{Appendix}

This Appendix contains (a) additional simulation results (Tables \ref{tab:SimResults1} - \ref{tab:SimResults6}), and (b) coefficients and standard errors for various fitted (penalized) models (Tables \ref{tab:LC_VSestimates1} and \ref{tab:LC_VSestimates2}).

\begin{table}[!h]
 \caption{Simulation results: variable selection metrics averaged over 200 simulation replicates.\label{tab:SimResults1}}	
 \begin{center}	
 \begin{adjustbox}{width=1\textwidth}
	\begin{threeparttable}
  {\footnotesize \begin{tabular}{cr|cccc|cccc|cccc}
             \hline
             \multicolumn{2}{c}{}&\multicolumn{12}{c}{One Tuning Parameter}\\
             \hline
             {} & {} & \multicolumn{4}{c|}{LASSO} &\multicolumn{4}{c|}{SCAD} &\multicolumn{4}{c}{ALASSO}\\
             $p_{\text{cen}}$ = 50\% & $n$ & C(0) & IC(7)& PT & MSE &C(7) & IC(0)&PT& MSE &C(7) & IC(0)&PT & MSE\\

             % {}&100& 5.29 & 0.18 & 0.12 & 0.33
             % &6.32 &  0.14 &  0.53 & 0.33
             % & 6.17 & 0.05 & 0.38 & 0.19  \\
             % Scale ($\beta$)&500 &5.52 & 0.00 & 0.18 & 0.09
             % &6.96 & 0.00 & 0.98 & 0.03
             % & 6.79 & 0.00 & 0.83 & 0.03\\
             % {} & 1000 & 5.83 & 0.00 & 0.31 & 0.05
             % &6.99 & 0.00  & 0.99 & 0.01
             % & 6.93 & 0.00 &0.94 & 0.01 \\
             % \hline
             % {}&100& 3.40 &  0.06 & 0.00 & 0.05
             % & 4.19 & 0.09 & 0.02 & 0.05
             % & 6.25 & 0.16 & 0.40 &  0.04 \\
             % Shape ($\alpha$)& 500 & 3.37 &  0.00 & 0.01 & 0.01
             % & 5.86  &0.00 & 0.33 & 0.01
             % &  6.81 & 0.00 & 0.85 & 0.00 \\
             % {} & 1000 & 3.46 & 0.00 & 0.00 & 0.01
             % &6.51 & 0.00 & 0.64 & 0.00
             % & 6.93 & 0.00 & 0.95 & 0.00 \\
             % \hline
             %$p_{\text{cen}}$ = 50\% & \multicolumn{13}{c}{\ } \\
             \hline
             {}&100& 5.29 & 0.34 &  0.08 & 0.46
             & 6.41 & 0.23 &  0.49 & 0.40
             & 6.29 & 0.17 & 0.39 & 0.30\\
             Scale ($\beta$)&500 & 5.65 & 0.00 & 0.23 & 0.10
             & 6.91 & 0.00 & 0.96 & 0.03
             & 6.66 & 0.00 & 0.78 & 0.03  \\
             {} & 1000 & 5.80 & 0.00 & 0.30 & 0.07
             & 6.95 & 0.00 & 0.98 & 0.02
             &6.85 & 0.00 & 0.95 & 0.02\\
             \hline	
             {}&100& 3.95 & 0.32 & 0.01 & 0.10
             & 5.23  & 0.45 &0.07 & 0.10
             & 6.19 & 0.52 & 0.27 & 0.11 \\
             Shape ($\alpha$)&500 & 4.02 & 0.00 & 0.03 & 0.01
             &6.19  & 0.02 & 0.45 & 0.01
             &  6.75 & 0.00 & 0.82 & 0.01 \\
             {} & 1000 & 4.27 & 0.00 & 0.04 & 0.01
             &6.62 & 0.00 & 0.70 & 0.00
             &6.79 & 0.00 & 0.91 & 0.00\\
             \hline
             \multicolumn{2}{c}{}&\multicolumn{12}{c}{Two Tuning Parameters}\\
             \hline
             {} & {} & \multicolumn{4}{c|}{LASSO} &\multicolumn{4}{c|}{SCAD}&\multicolumn{4}{c}{ALASSO}\\
             $p_{\text{cen}}$ = 50\% & $n$ & C(7) & IC(0) & PT & MSE &C(7) & IC(0)&PT& MSE &C(7) & IC(0)&PT & MSE\\
             \hline
             % {}&100& 4.61 & 0.10 & 0.06 & 0.28
             % & 6.26 & 0.10 & 0.55 & 0.28
             % & 6.47 & 0.14 & 0.52 & 0.23 \\
             % Scale ($\beta$)&500 & 5.40 & 0.00 & 0.20 & 0.08
             % & 6.95 & 0.00& 0.97 & 0.02
             % & 6.84 & 0.00 &0.88 & 0.03 \\
             % {} & 1000 & 5.32 & 0.00 & 0.20 & 0.04 & 6.89& 0.00& 0.96& 0.01 &6.92 & 0.00& 0.92 & 0.01\\
             % \hline
             % {}&100& 5.27 & 0.25 &0.10 & 0.05
             % & 5.15 & 0.20 & 0.11 & 0.06
             % &6.30 & 0.25 & 0.44 & 0.05\\
             % Shape ($\alpha$)&500 & 5.30 & 0.00 & 0.17 & 0.01
             % & 5.98 & 0.00 & 0.36 & 0.01
             % &6.93 & 0.00 & 0.93 & 0.00\\
             % {} & 1000 & 5.50 & 0.00& 0.26 & 0.01 & 6.51& 0.00 &0.63 & 0.00 & 6.94 &0.00 & 0.94 & 0.00\\
             % \hline
             %$p_{\text{cen}}$ = 50\% & \multicolumn{13}{c}{\ } \\
             %\hline
             {}&100& 4.95 & 0.26 & 0.09 & 0.45
             & 6.36 & 0.26 & 0.47 & 0.40
             & 6.18 & 0.35 & 0.41 & 0.33\\
             Scale ($\beta$)&500 & 5.47 & 0.00 & 0.19 & 0.08
             & 6.92 & 0.00 & 0.96 & 0.03
             & 6.85 & 0.00 & 0.87 & 0.03 \\
             {} & 1000 & 5.56 & 0.00& 0.21 & 0.05 & 6.86 & 0.00 & 0.96& 0.01& 6.95 & 0.00& 0.95 & 0.02\\
             \hline	
             {}&100& 5.72 & 0.74 & 0.05 & 0.11
             &5.58 & 0.61 & 0.08 & 0.12
             &6.28 &0.54 & 0.23 &  0.08 \\
             Shape($\alpha$)&500 & 5.58 & 0.02 & 0.26 & 0.01
             & 6.27 & 0.01 & 0.52 & 0.01
             &6.84 &0.01 & 0.87 & 0.01 \\
             {} & 1000 & 5.62 & 0.00 & 0.24 & 0.01 & 6.71& 0.00&0.77 &0.00 &6.93 & 0.00& 0.93 &0.00\\
             \hline
		 \end{tabular}}
		 {\footnotesize\begin{tablenotes}
			\item[]{ {C, average correct zeros; IC, average incorrect zeros; PT, the probability of choosing the true model; MSE, the average mean squared error.}}
			\end{tablenotes}}
			\end{threeparttable}
\end{adjustbox}
\end{center}
\end{table}

\begin{table}%[!ht] \centering \footnotesize
   \caption{Further simulation results: estimates, standard errors, and confidence intervals.}
   \begin{center}
	\begin{threeparttable}
   \begin{adjustbox}{width=1\textwidth}
			{\footnotesize\begin{tabular}{cc|cccc|cccc|cccc}
				\hline
				\multicolumn{14}{c}{ALASSO}\\
				\hline
				\multicolumn{14}{c}{One Tuning Parameter}\\
				\hline
				$p_{\text{cen}}$= 50\%& & \multicolumn{4}{c|}{$n$ = 100} & \multicolumn{4}{c|}{$n$ = 500} & \multicolumn{4}{c}{$n$ = 1000}\\
				{} & $\theta$ &$\hat{\theta}$ & SE & SEE & CP & $\hat{\theta}$ & SE & SEE & CP& $\hat{\theta}$ & SE & SEE & CP\\
				\hline
				$\beta_0$ & -1.50  &-1.47 & 0.28 & 0.23 & 0.90
				& -1.48 & 0.10 & 0.10 & 0.95
				& -1.49 & 0.08 & 0.07& 0.93\\
				$\beta_1$ & -1.00 & -0.96 & 0.23 & 0.18 & 0.88
				& -0.99 & 0.07 & 0.08 & 0.97
				&  -0.99 & 0.06 & 0.05& 0.94\\
				$\beta_7$ & -0.80&  -0.72 & 0.23 & 0.18 & 0.78
				& -0.76 & 0.08 & 0.07 & 0.92
				& -0.79 & 0.06 & 0.05& 0.90\\
				$\beta_8$ & 0.50 & 0.36 & 0.24 & 0.14 & 0.76
				& 0.45 & 0.08 & 0.07 & 0.87
				& 0.48 & 0.05 & 0.05 & 0.95\\
				\hline
				$\alpha_0$ & 0.50 & 0.54 & 0.14& 0.12& 0.86
				& 0.51& 0.05& 0.05& 0.94
				& 0.51& 0.04& 0.04& 0.94\\
				$\alpha_1$ & 0.40 & 0.35 & 0.11 & 0.07& 0.85
				& 0.40& 0.03& 0.03& 0.97
				& 0.40& 0.02& 0.02& 0.96 \\
				$\alpha_5$ & 0.40 &0.29 & 0.12 & 0.08& 0.70
				& 0.38& 0.04 & 0.04& 0.91
				& 0.39& 0.03& 0.03& 0.91\\
				$\alpha_6$ &-0.20 & -0.09 & 0.10 & 0.05& 0.53
				& -0.17& 0.04 & 0.04 & 0.82
				& -0.18& 0.03 & 0.03& 0.86\\
				\hline
				\multicolumn{14}{c}{Two Tuning Parameters}\\
				\hline
				$p_{\text{cen}}$= 50\% & & \multicolumn{4}{c|}{$n$ = 100} & \multicolumn{4}{c|}{$n$ = 500}& \multicolumn{4}{c}{$n$ = 1000}\\
				{} & $\theta$ &$\hat{\theta}$ & SE & SEE & CP & $\hat{\theta}$ & SE & SEE & CP& $\hat{\theta}$ & SE & SEE & CP\\
				\hline
				$\beta_0$ & -1.50  &-1.48 & 0.27 & 0.23 & 0.90
				& -1.47 & 0.11 & 0.10 & 0.95
				& -1.50 & 0.07 & 0.07& 0.94\\
				$\beta_1$ & -1.00 & -1.00 & 0.25 & 0.19 & 0.88
				& -0.98 & 0.08 & 0.07 & 0.97
				&  -1.00 & 0.06 & 0.05& 0.93\\
				$\beta_7$ & -0.80&  -0.73 & 0.26 & 0.17 & 0.78
				& -0.77 & 0.08 & 0.07 & 0.92
				& -0.78 & 0.05 & 0.05& 0.95\\
				$\beta_8$ & 0.50 & 0.36 & 0.25 & 0.13 & 0.76
				& 0.47 & 0.07 & 0.07 & 0.87
				& 0.48 & 0.04 & 0.05 & 0.97\\
				\hline
				$\alpha_0$ & 0.50 & 0.54 & 0.14& 0.12& 0.90
				& 0.51& 0.05& 0.05& 0.95
				& 0.51& 0.04& 0.04& 0.95\\
				$\alpha_1$ & 0.40 & 0.37 & 0.11 & 0.07& 0.84
				& 0.39& 0.03& 0.03& 0.93
				& 0.40& 0.02& 0.02& 0.95\\
				$\alpha_5$ & 0.40 &0.32 & 0.13 & 0.08& 0.73
				& 0.38& 0.04 & 0.04& 0.93
				& 0.39& 0.03& 0.03& 0.91\\
				$\alpha_6$ &-0.20 & -0.11 & 0.12 & 0.05& 0.51
				& -0.18& 0.04 & 0.04 & 0.87
				& -0.19& 0.03 & 0.02& 0.91\\
				\hline
   \end{tabular}}
 \end{adjustbox}
{\footnotesize\begin{tablenotes}
	\item[]{ SE, standard deviation of estimates over 200 replications; SEE, average of estimated standard errors over 200 replications; CP, the empirical coverage probability of a nominal 95\% confidence interval.}
\end{tablenotes}}
\end{threeparttable}
\end{center}
\end{table}

		\begin{table} %[!ht] \centering \footnotesize
   \caption{Further simulation results: estimates, standard errors, and confidence intervals.}
   \begin{center}
   \begin{adjustbox}{width=1\textwidth}
			{\footnotesize\begin{tabular}{cc|cccc|cccc|cccc}
				\hline
				\multicolumn{14}{c}{LASSO}\\
				\hline
				\multicolumn{14}{c}{One Tuning Parameter}\\
				\hline
				\bf{$p_{\text{cen}}$= 25\% }& & \multicolumn{4}{c|}{$n$ = 100} & \multicolumn{4}{c|}{$n$ = 500}& \multicolumn{4}{c}{$n$ = 1000}\\
				{} & $\theta$ & $\hat{\theta}$ & SE & SEE & CP & $\hat{\theta}$ & SE & SEE & CP & $\hat{\theta}$ & SE & SEE & CP\\
				\hline
				$\beta_0$ & -1.50 & -1.39 & 0.25 & 0.20 & 0.80
				& -1.37 & 0.10 & 0.09 & 0. 67
				& -1.39 & 0.07 & 0.06 & 0.55\\
				$\beta_1$ & -1.00 & -0.82 & 0.21 & 0.17 & 0.74
				& -0.88 & 0.08 & 0.07 & 0.62
				&  -0.91 & 0.06 & 0.05 & 0.49\\
				$\beta_7$ & -0.80&  -0.55 & 0.21 & 0.16 & 0.65
				& -0.66 & 0.08 & 0.07 & 0.48
				& -0.69 & 0.05 & 0.05 & 0.35\\
				$\beta_8$ & 0.50 & 0.25 & 0.18 & 0.13 & 0.62
				& 0.37 & 0.07 & 0.07 & 0.48
				& 0.40 & 0.05 & 0.05 & 0.37\\
				% Average across &  \multicolumn{13}{c}{} \\
				% zero-coefficients & 0.00& 0.00& 0.06&0.03 & 1 &0.00 &0.02 & 0.01 &1 &0.00 & 0.02& 0.01 & 1 \\
				\hline
				$\alpha_0$ & 0.50 & 0.49 & 0.12 & 0.10 &  0.89
				&0.46 &  0.05 & 0.04 & 0.82
				& 0.46& 0.03& 0.03& 0.72\\
				$\alpha_1$ & 0.40 & 0.35 & 0.08 & 0.07 & 0.86
				& 0.38 & 0.03 & 0.03 & 0.85
				& 0.38& 0.02& 0.02& 0.84\\
				$\alpha_5$ & 0.40 & 0.35 & 0.09 & 0.07 & 0.87
				& 0.38& 0.03 & 0.03& 0.89
				& 0.39& 0.02& 0.02& 0.92\\
				$\alpha_6$ &-0.20 & -0.16 &0.08 &0.07 &0.90
				& -0.18 & 0.03 & 0.03& 0.91
				& -0.19& 0.02& 0.02& 0.91 \\
				% Average across & \multicolumn{13}{c}{} \\
				% zero-coefficients & 0.00& 0.00& 0.05&0.03 & 0.97 &0.00& 0.02&0.01 & 0.98 &0.00& 0.01& 0.01& 0.98 \\
				\hline
				\multicolumn{14}{c}{Two Tuning Parameters}\\
				\hline
				$p_{\text{cen}}$= 25\%& & \multicolumn{4}{c|}{$n$ = 100}& \multicolumn{4}{c|}{$n$ = 500} & \multicolumn{4}{c}{$n$ = 1000} \\
				{} & $\theta$ & $\hat{\theta}$ & SE & SEE & CP & $\hat{\theta}$ & SE & SEE & CP & $\hat{\theta}$ & SE & SEE & CP\\
				\hline
				$\beta_0$ & -1.50 & -1.39 & 0.22 & 0.20 & 0.88
				& -1.39 & 0.10 & 0.09 & 0.68
				& -1.41 & 0.07& 0.06 & 0.61\\
				$\beta_1$ & -1.00 & -0.84 & 0.22 & 0.17 & 0.77
				& -0.89 & 0.08 & 0.07 & 0.60
				&  -0.92 & 0.05 & 0.05 & 0.59\\
				$\beta_7$ & -0.80&  -0.67 & 0.20 & 0.16 & 0.80
				& -0.69 & 0.08 & 0.06 & 0.62
				& -0.72 & 0.06 & 0.05 & 0.56\\
				$\beta_8$ & 0.50 & 0.36 & 0.19 & 0.13 & 0.78
				& 0.41 & 0.08 & 0.06 & 0.63
				& 0.43 & 0.05& 0.04& 0.56 \\
				% Average across & \multicolumn{13}{c}{}\\
				% zero-coefficients &0.00 &0.00 &0.1 &0.05 &0.97 &0.00 &0.03 & 0.01&0.99 & 0.00& 0.02& 0.01 &0.99 \\
				\hline
				$\alpha_0$ & 0.50 & 0.50 & 0.12 & 0.10 &  0.93
				&0.47 &  0.05& 0.04 & 0.81
				& 0.47& 0.03& 0.03& 0.82\\
				$\alpha_1$ & 0.40 & 0.34 & 0.08 & 0.06& 0.83
				& 0.37 & 0.02 & 0.02 & 0.80
				& 0.38& 0.02& 0.02& 0.78\\
				$\alpha_5$ & 0.40 & 0.32 & 0.09 & 0.07& 0.73
				& 0.36& 0.03 & 0.03& 0.72
				& 0.37& 0.02& 0.02& 0.66\\
				$\alpha_6$ &-0.20 & -0.11 &0.09 &0.05 &0.69
				& -0.16 & 0.03 & 0.03& 0.69
				& -0.17& 0.02& 0.02& 0.61 \\
				% Average across & \multicolumn{13}{c}{} \\
				% zero-coefficients &0.00 &0.00 &0.04 &0.02 & 0.98&0.00 &0.01 & 0.01&0.99 &0.00 & 0.01&0.00 & 0.99\\
				\hline
   \end{tabular}}
 \end{adjustbox}
{\footnotesize\begin{tablenotes}
	\item[]{ SE, standard deviation of estimates over 200 replications; SEE, average of estimated standard errors over 200 replications; CP, the empirical coverage probability of a nominal 95\% confidence interval.}
\end{tablenotes}}
\end{center}
\end{table}

\begin{table} %[!ht] \centering \footnotesize
   \caption{Further simulation results: estimates, standard errors, and confidence intervals.}
   \begin{center}
   \begin{adjustbox}{width=1\textwidth}
			{\footnotesize\begin{tabular}{cc|cccc|cccc|cccc}
				\hline
				\multicolumn{14}{c}{LASSO}\\
				\hline
				\multicolumn{14}{c}{One Tuning Parameter}\\
				\hline
				$p_{\text{cen}}$= 50\% & & \multicolumn{4}{c|}{$n$ = 100}& \multicolumn{4}{c|}{$n$ = 500} & \multicolumn{4}{c}{$n$ = 1000} \\
				{} & $\theta$ &$\hat{\theta}$ & SE & SEE & CP & $\hat{\theta}$ & SE & SEE & CP& $\hat{\theta}$ & SE & SEE & CP\\
				\hline
				$\beta_0$ & -1.50  &-1.37 & 0.26 & 0.21 & 0.76
				& -1.35 & 0.10 & 0.09 & 0.62
				& -1.37 & 0.07 & 0.07& 0.52\\
				$\beta_1$ & -1.00 & -0.77 & 0.27 & 0.18 & 0.65
				& -0.87 & 0.08 & 0.08 & 0.59
				&  -0.91 & 0.06 & 0.05& 0.49\\
				$\beta_7$ & -0.80&  -0.51 & 0.24 & 0.18 & 0.55
				& -0.64 & 0.08 & 0.08 & 0.43
				& -0.69 & 0.06 & 0.05& 0.34\\
				$\beta_8$ & 0.50 & 0.21 & 0.20 & 0.12 & 0.53
				& 0.35 & 0.07 & 0.07 & 0.44
				& 0.40 & 0.06 & 0.05 & 0.37\\
				% Average across & \multicolumn{13}{c}{} \\
				% zero-coefficients & 0.00 & -0.01& 0.08 & 0.04 & 0.99& 0.00& 0.02 & 0.01& 1& 0.00 &0.02 & 0.01 & 0.99\\
				\hline
				$\alpha_0$ & 0.50 & 0.47 & 0.14& 0.12& 0.99
				& 0.46& 0.05& 0.05& 0.86
				& 0.46& 0.04& 0.04& 0.78\\
				$\alpha_1$ & 0.40 & 0.32 & 0.11 & 0.08& 0.82
				& 0.38& 0.03& 0.03& 0.90
				& 0.38& 0.02& 0.02& 0.85 \\
				$\alpha_5$ & 0.40 &0.31 & 0.11 & 0.10& 0.80
				& 0.36& 0.04 & 0.04& 0.88
				& 0.38& 0.03& 0.03& 0.86\\
				$\alpha_6$ &-0.20 & -0.12 & 0.11 & 0.07& 0.67
				& -0.16& 0.04 & 0.04 & 0.86
				& -0.17& 0.03 & 0.03& 0.85\\
				% Average across & \multicolumn{13}{c}{} \\
				% zero-coefficients & 0.00& 0.00&0.07 & 0.04& 0.98 &0.00&0.02  & 0.02& 0.99 & 0.00 & 0.01& 0.01 & 0.99 \\
				
				\hline
				\multicolumn{14}{c}{Two Tuning Parameters}\\
				\hline
				$p_{\text{cen}}$= 50\% & & \multicolumn{4}{c|}{$n$ = 100}& \multicolumn{4}{c|}{$n$ = 500}& \multicolumn{4}{c}{$n$ = 1000}\\
				{} & $\theta$ &$\hat{\theta}$ & SE & SEE & CP & $\hat{\theta}$ & SE & SEE & CP& $\hat{\theta}$ & SE & SEE & CP\\
				\hline
				$\beta_0$ & -1.50  &-1.31 & 0.29 & 0.23 & 0.71
				& -1.38 & 0.10 & 0.09 & 0.73
				& -1.40 & 0.07 & 0.07& 0.65\\
				$\beta_1$ & -1.00 & -0.81 & 0.24 & 0.18 & 0.75
				& -0.89 & 0.09 & 0.07 & 0.68
				&  -0.91 & 0.05 & 0.05& 0.64\\
				$\beta_7$ & -0.80&  -0.57 & 0.26 & 0.18 & 0.65
				& -0.68 & 0.08 & 0.07 & 0.64
				& -0.70 & 0.06 & 0.05& 0.56\\
				$\beta_8$ & 0.50 & 0.28 & 0.23 & 0.14 & 0.65
				& 0.39 & 0.08 & 0.06 & 0. 66
				& 0.40 & 0.05 & 0.05 & 0.49\\
				% Average across & \multicolumn{13}{c}{} \\
				% zero-coefficients &0.00 &0.00 & 0.11&0.04 &0.97 & 0.00& 0.03&0.01 &0.99 & 0.00&0.02 & 0.01& 0.99\\
				\hline
				$\alpha_0$ & 0.50 & 0.48 & 0.15& 0.12& 0.90
				& 0.48& 0.05& 0.05& 0.93
				& 0.48& 0.04& 0.04& 0.87\\
				$\alpha_1$ & 0.40 & 0.31 & 0.10 & 0.08& 0.79
				& 0.37& 0.03& 0.03& 0.86
				& 0.38& 0.02& 0.02& 0.85 \\
				$\alpha_5$ & 0.40 &0.24 & 0.13 & 0.09& 0.54
				& 0.35& 0.04 & 0.04& 0.74
				& 0.36& 0.03& 0.03& 0.61\\
				$\alpha_6$ &-0.20 & -0.06 & 0.09 & 0.03& 0.35
				& -0.14& 0.04 & 0.04 & 0.66
				& -0.15& 0.03& 0.03& 0.62\\
				% Average across & \multicolumn{13}{c}{} \\
				% zero-coefficients & 0.00& 0.01 &0.05 &0.02 &0.98 &0.00 &0.02 & 0.01&0.99 &0.00 &0.01 &0.01 & 0.99\\
				\hline
   \end{tabular}}
 \end{adjustbox}
 {\footnotesize\begin{tablenotes}
	\item[]{ SE, standard deviation of estimates over 200 replications; SEE, average of estimated standard errors over 200 replications; CP, the empirical coverage probability of a nominal 95\% confidence interval.}
\end{tablenotes}}
\end{center}
\end{table}

\begin{table}%[!ht] \centering \footnotesize
   \caption{Further simulation results: estimates, standard errors, and confidence intervals.}
   \begin{center}
   \begin{adjustbox}{width=1\textwidth}
     {\footnotesize\begin{tabular}{cc|cccc|cccc|cccc}
				\hline
				\multicolumn{14}{c}{SCAD}\\
				\hline
				\multicolumn{14}{c}{One Tuning Parameter}\\
				\hline
				$p_{\text{cen}}$ = 25\%& & \multicolumn{4}{c|}{$n$ = 100} & \multicolumn{4}{c|}{$n$ = 500}& \multicolumn{4}{c}{$n$ = 1000}\\
				{} & $\theta$ & $\hat{\theta}$ & SE & SEE & CP & $\hat{\theta}$ & SE & SEE & CP & $\hat{\theta}$ & SE & SEE & CP\\
				\hline
				$\beta_0$ & -1.50 & -1.62 & 0.28 & 0.22 & 0.86
				& -1.54 & 0.10 & 0.09 & 0.90
				& -1.50 & 0.06 & 0.06 & 0.96\\
				$\beta_1$ & -1.00 & -1.07 & 0.23 & 0.17 & 0.84
				& -1.01 & 0.07 & 0.07& 0.91
				&  -1.00 & 0.04 & 0.05 & 0.98\\
				$\beta_7$ & -0.80&  -0.87 & 0.25 & 0.15 & 0.76
				& -0.83 & 0.07 & 0.06 & 0.95
				& -0.81 & 0.05 & 0.04 & 0.92\\
				$\beta_8$ & 0.50 & 0.50 & 0.27 & 0.11 & 0.59
				& 0.51 & 0.06 & 0.05& 0.82
				& 0.51 & 0.04 & 0.04  & 0.88 \\
				% Average across & \multicolumn{13}{c}{} \\
				% zero-coefficients &0.00 &0.00 &0.08 &0.01 & 0.95 &0.00 &0.00 &0.00 &1 &0.00 & 0.00& 0.00 &1 \\
				\hline
				$\alpha_0$ & 0.50 & 0.60 & 0.12 & 0.10&  0.77
				&0.52 &  0.04& 0.04 & 0.92
				& 0.51 & 0.03& 0.03& 0.95\\
				$\alpha_1$ & 0.40 & 0.39 & 0.08 & 0.06& 0.80
				& 0.40 & 0.02 & 0.02 & 0.79
				& 0.40 & 0.02& 0.01& 0.80\\
				$\alpha_5$ & 0.40 & 0.36 & 0.09 & 0.07 & 0.74
				& 0.38& 0.03 & 0.02 & 0.63
				& 0.39& 0.02& 0.01& 0.75\\
				$\alpha_6$ &-0.20 & -0.15 &0.09 &0.06 &0.69
				& -0.17 & 0.03 & 0.02 & 0.77
				& -0.18 & 0.02& 0.02 & 0.75 \\
				% Average across &\multicolumn{13}{c}{} \\
				% zero-coefficients & 0.00 & 0.00 & 0.04 & 0.03 &0.96 &0.00 &0.01 &0.00 &1 &0.00 &0.00 & 0.00 & 1\\

				\hline
				\multicolumn{14}{c}{Two Tuning Parameters}\\
				\hline
				$p_{\text{cen}}$ = 25\%& & \multicolumn{4}{c|}{$n$ = 100} & \multicolumn{4}{c|}{$n$ = 500}& \multicolumn{4}{c}{$n$ = 1000}\\
				{} & $\theta$ & $\hat{\theta}$ & SE & SEE & CP & $\hat{\theta}$ & SE & SEE & CP & $\hat{\theta}$ & SE & SEE & CP\\
				\hline
				$\beta_0$ & -1.50 & -1.59 &0.26 & 0.22 &0.91
				&-1.53 & 0.09& 0.09& 0.96
				&-1.51 &0.06 & 0.06 & 0.94\\
				$\beta_1$ & -1.00 &-1.05 & 0.22& 0.17 &0.89
				& -1.01& 0.07& 0.07&0.95
				& -1.00& 0.05& 0.05& 0.96\\
				$\beta_7$ & -0.80& -0.86 & 0.23& 0.16& 0.84
				& -0.81& 0.07& 0.06&0.88
				& -0.81& 0.04& 0.04& 0.90\\
				$\beta_8$ & 0.50 & 0.51 & 0.22 & 0.10& 0.70
				& 0.50& 0.06& 0.02& 0.55
				& 0.50& 0.04 & 0.01 & 0.96 \\
				% Average across & \multicolumn{13}{c}{}\\
				% zero-coefficients & 0.00 & 0.01& 0.1& 0.01& 0.94 &0.00 & 0.00& 0.00& 1 &0.00 &0.01 & 0.00&0.99 \\
				\hline
				$\alpha_0$ & 0.50 & 0.58 & 0.11 & 0.10& 0.79
				& 0.52& 0.04& 0.04& 0.91
				& 0.51& 0.03 &0.03 & 0.93\\
				$\alpha_1$ & 0.40 & 0.37& 0.08& 0.06& 0.83
				& 0.40& 0.02& 0.02 & 0.78
				&0.40 & 0.01 & 0.01& 0.81\\
				$\alpha_5$ & 0.40 & 0.34& 0.10& 0.06& 0.68
				& 0.38& 0.03& 0.02&0.64
				&0.39 & 0.02& 0.01 & 0.60\\
				$\alpha_6$ &-0.20 &-0.12 & 0.06& 0.05& 0.68
				& -0.17& 0.03& 0.02& 0.64
				& -0.18& 0.02& 0.02 &0.80\\
				% Average across & \multicolumn{13}{c}{} \\
				% zero-coefficients & 0.00& 0.00& 0.04 &0.02 & 0.96&0.00 & 0.01& 0.00& 0.99& 0.00& 0.00& 0.00&1 \\
				\hline
   \end{tabular}}
 \end{adjustbox}
{\footnotesize\begin{tablenotes}
	\item[]{ SE, standard deviation of estimates over 200 replications; SEE, average of estimated standard errors over 200 replications; CP, the empirical coverage probability of a nominal 95\% confidence interval.}
\end{tablenotes}}
\end{center}
\end{table}

\begin{table}%[!ht] \centering \footnotesize
   \caption{Further simulation results: estimates, standard errors, and confidence intervals.\label{tab:SimResults6}}
   \begin{center}
   \begin{adjustbox}{width=1\textwidth}
			{\footnotesize\begin{tabular}{cc|cccc|cccc|cccc}
				\hline
				\multicolumn{14}{c}{SCAD}\\
				\hline
				\multicolumn{14}{c}{One Tuning Parameter}\\
				\hline
				$p_{\text{cen}}$ = 50\% & & \multicolumn{4}{c|}{$n$ = 100}& \multicolumn{4}{c|}{$n$ = 500} & \multicolumn{4}{c}{$n$ = 1000}\\
				{} & $\theta$ &$\hat{\theta}$ & SE & SEE & CP & $\hat{\theta}$ & SE & SEE & CP& $\hat{\theta}$ & SE & SEE & CP\\
				\hline
				$\beta_0$ & -1.50  &-1.60 & 0.30 & 0.25 & 0.89
				& -1.52 & 0.10 & 0.10 & 0.97
				& -1.51 & 0.07 & 0.07& 0.94\\
				$\beta_1$ & -1.00 & -1.07 & 0.23 & 0.19 & 0.90
				& -1.01 & 0.08 & 0.07 & 0.96
				&  -1.01 & 0.05 & 0.05& 0.95\\
				$\beta_7$ & -0.80&  -0.84 & 0.30 & 0.18& 0.84
				& -0.81 & 0.08 & 0.07 & 0.96
				& -0.81 & 0.06 & 0.05 & 0.92\\
				$\beta_8$ & 0.50 & 0.50 & 0.31 & 0.12 & 0.58
				& 0.51 & 0.07 & 0.07 & 0.94
				& 0.50 & 0.05 & 0.05 & 0.94\\
				% Average across & \multicolumn{13}{c}{} \\
				% zero-coefficients & 0.00 &0.00 &0.11 &0.01 &0.95 &0.00 &0.01 &0.00 &1 &0.00 &0.00 & 0.00 & 1\\
				\hline
				$\alpha_0$ & 0.50 & 0.59 & 0.14 & 0.12 & 0.84
				& 0.53& 0.05& 0.05& 0.89
				& 0.51& 0.04& 0.04& 0.93\\
				$\alpha_1$ & 0.40 & 0.37 & 0.12 & 0.08& 0.74
				& 0.40& 0.03& 0.03 & 0.90
				& 0.40& 0.02& 0.02& 0.96 \\
				$\alpha_5$ & 0.40 &0.35 & 0.16 & 0.08& 0.64
				& 0.38& 0.05 & 0.03 & 0.63
				& 0.39& 0.03& 0.02& 0.82\\
				$\alpha_6$ &-0.20 & -0.13 & 0.15 & 0.05& 0.41
				& -0.16& 0.05 & 0.03 & 0.65
				& -0.18& 0.04 & 0.03 & 0.78\\
				% Average across & \multicolumn{13}{c}{} \\
				% zero-coefficients &0.00 &0.00 &0.06 &0.02 &0.96 &0.00 &0.01 &0.00 &0.99 & 0.00 & 0.01 & 0.00 & 1\\
				
				\hline
				\multicolumn{14}{c}{Two Tuning Parameters}\\
				\hline
				$p_{\text{cen}}$ = 50\% & & \multicolumn{4}{c|}{$n$ = 100} & \multicolumn{4}{c|}{$n$ = 500}& \multicolumn{4}{c}{$n$ = 1000}\\
				{} & $\theta$ &$\hat{\theta}$ & SE & SEE & CP & $\hat{\theta}$ & SE & SEE & CP& $\hat{\theta}$ & SE & SEE & CP\\
				\hline
				$\beta_0$ & -1.50  & -1.61& 0.30& 0.24& 0.86
				& -1.52 & 0.09& 0.10& 0.97
				& -1.50&0.07 &0.07 & 0.96\\
				$\beta_1$ & -1.00 & -1.07& 0.21& 0.19 & 0.90
				&-1.02 & 0.07& 0.07& 0.96
				& -1.01& 0.05 & 0.05& 0.95\\
				$\beta_7$ & -0.80& -0.85 & 0.27 & 0.16& 0.78
				&-0.81 & 0.08& 0.07& 0.94
				& -0.80& 0.05 & 0.05& 0.92\\
				$\beta_8$ & 0.50 & 0.48& 0.30& 0.10& 0.52
				& 0.49& 0.07& 0.03 & 0.55
				& 0.50& 0.05& 0.01 & 0.92 \\
				% Average across & \multicolumn{13}{c}{} \\
				% zero-coefficients & 0.00 &0.00 & 0.12 & 0.01& 0.93&0.00 &0.01 &0.00 & 1 &0.00 & 0.01& 0.00&1 \\
				\hline
				$\alpha_0$ & 0.50 & 0.60& 0.15& 0.12 & 0.79
				& 0.52& 0.05& 0.05& 0.92
				& 0.51& 0.04& 0.03&0.93\\
				$\alpha_1$ & 0.40& 0.36& 0.13 & 0.07& 0.75
				& 0.40& 0.03& 0.03 & 0.88
				& 0.40& 0.02& 0.02& 0.92\\
				$\alpha_5$ & 0.40 & 0.30& 0.16& 0.07& 0.49
				&0.38 & 0.04& 0.03& 0.68
				&0.39 & 0.03& 0.02 &0.79\\
				$\alpha_6$ &-0.20 & -0.11& 0.14& 0.04& 0.34
				& -0.16 & 0.05& 0.04& 0.66
				& -0.18& 0.04& 0.02& 0.70\\
				% Average across & \multicolumn{13}{c}{} \\
				% zero-coefficients & 0.00 & 0.00& 0.06 &0.02 & 0.95&0.00 & 0.01 & 0.00& 0.99&0.00 & 0.00& 0.00& 1 \\
				\hline
   \end{tabular}}
 \end{adjustbox}
{\footnotesize\begin{tablenotes}
	\item[]{ SE, standard deviation of estimates over 200 replications; SEE, average of estimated standard errors over 200 replications; CP, the empirical coverage probability of a nominal 95\% confidence interval.}
	\end{tablenotes}}
\end{center}
\end{table}

\begin{sidewaystable}
		\caption{Coefficients estimates and standard errors for the one tuning parameter setup {(lung cancer dataset)} \label{tab:LC_VSestimates1}}\centering
		\begin{threeparttable}
		\begin{adjustbox}{max width=\linewidth}
		{\footnotesize\begin{tabular}{ll|cccc|cccc}
			\hline
			{} & {} & \multicolumn{4}{c|}{Scale}& \multicolumn{4}{c}{Shape} \\
			\hline
			\multicolumn{2}{c|}{Covariate}& No Penalt & LASSO & SCAD & ALASSO & No Penalty & LASSO & SCAD & ALASSO\\
			Intercept & & \bf{-3.38 (0.66)} & \bf{-2.66 (0.25)} & \bf{-3.35 (0.21)} & \bf{-3.12 (0.17)} & -0.16 (0.22) & -0.13 (0.09) & -0.05 (0.08) & 0.04 (0.03) \\
			Treatment & surgery & \bf{-1.69 (0.83)}& -0.82 (0.50)& \bf{-1.21 (0.26)} & \bf{-0.89 (0.21)}  & 0.11 (0.21) & -0.14 (0.19) & 0.00 (0.00) & 0.00 (0.00) \\
			& chemotherapy &  -0.33 (0.37) & 0.00 (0.00) & 0.00  (0.00) & 0.00  (0.00)  & -0.03 (0.15) & -0.14 (0.09) & -0.09 (0.08) & 0.00 (0.00) \\
			& radiotherapy & \bf{-0.85 (0.21)} & \bf{-0.44 (0.19)} & \bf{-0.84 (0.26)} & -0.16 (0.10) & \bf{0.22 (0.08) }& 0.09 (0.08) & \bf{0.22 (0.10)} & 0.00 (0.00) \\
			& chemo. \& radio. & \bf{-3.83 (0.98)}& -0.85 (0.59) & \bf{-3.92 (0.93)} & \bf{-2.30 (0.89)} & \bf{0.77 (0.20)} & 0.06 (0.20) & \bf{0.82 (0.17)} & \bf{0.51 (0.21)} \\
			Age group & $50-$ &  \bf{-0.90 (0.43)} & 0.00 (0.00) &  0.00 (0.00) & 0.00 (0.00) &  \bf{0.39 (0.16)} & 0.05 (0.06) & 0.03 (0.05) & 0.00 (0.00)\\
			& $60-$ & \bf{-0.94 (0.39)} &0.00 (0.00) &  0.00 (0.00) &  0.00 (0.00) & \bf{0.40 (0.15)} & 0.05 (0.04) & 0.03 (0.03) & 0.00 (0.00) \\
			& $70-$ & -0.77 (0.39) & 0.00 (0.00) &  0.00 (0.00) & 0.02 (0.08) & \bf{0.31 (0.15)} & 0.00 (0.00) & 0.00 (0.00) & 0.00 (0.00) \\
			& $> 80$ & -0.78 (0.42) & 0.00 (0.00) &  0.00 (0.00) & 0.00 (0.00)  & 0.31 (0.17) & 0.00 (0.00) & 0.00 (0.00) & 0.00 (0.00)\\
			WHO status & light work & -0.02 (0.45) &-0.42 (0.27) & 0.00 (0.00) & 0.00 (0.00) & 0.02 (0.12) & 0.13 (0.08) &  \bf{0.09 (0.00)} & 0.00 (0.00) \\
			& unable to work & 0.84 (0.43) & 0.23 (0.23) & \bf{0.71 (0.10)} & \bf{0.41 (0.10)} & -0.10 (0.13) & 0.03 (0.07) & 0.00 (0.00) & 0.00 (0.00) \\ 	
			& $> 50\%$ walking & \bf{1.31 (0.44)} & \bf{0.79 (0.18)} & \bf{ 1.24 (0.17)} & \bf{0.99 (0.11)} & -0.13 (0.14) & 0.00 (0.00) & -0.02 (0.07) & 0.00 (0.00) \\
			& bed/chair bound & \bf{1.78 (0.50)} & \bf{1.27 (0.29)} &  \bf{1.80 (0.24)} & \bf{1.28 (0.28)} & -0.03 (0.20) & 0.00 (0.00) & 0.00 (0.00) & 0.00 (0.00) \\
			Sex & male & 0.03 (0.14) & 0.00 (0.00) & 0.00 (0.00) & 0.00 (0.00) & -0.03 (0.05) & -0.01 (0.04) & 0.00 (0.00) & 0.00 (0.00) \\
			Smoking status & current smoker & 0.10 (0.22) & 0.00 (0.00) & 0.00 (0.00) & 0.00 (0.00) & 0.15 (0.08) & 0.09 (0.05) & 0.08 (0.05) & 0.00 (0.00) \\
			& ex-smoker & -0.05 (0.23) & 0.00 (0.00) & 0.00 (0.00) & 0.00 (0.00) &  0.17 (0.09) & 0.05 (0.05) & 0.06 (0.05) & 0.00 (0.00) \\
			& missing &  0.29 (0.40) & 0.00 (0.00) & 0.00 (0.00) & 0.00 (0.00) & 0.00 (0.00) & 0.00 (0.00) & 0.00 (0.00) & 0.00 (0.00) \\
			Cell type & small cell &  \bf{0.83 (0.26)} & 0.23 (0.21) &  \bf{0.46 (0.15)} &  \bf{0.31 (0.12)} & -0.05 (0.10) & 0.11 (0.09) & 0.00 (0.00) & 0.00 (0.00) \\
			& adenocarcinoma & 0.28 (0.28) & 0.00 (0.00) & 0.00 (0.00) & 0.00 (0.00) & 0.03 (0.10) & 0.09 (0.05) & 0.06 (0.05) & 0.00 (0.00) \\
			& other & 0.32 (0.20) & 0.13 (0.10) & 0.03 (0.10) & 0.00 (0.00) & -0.04 (0.07) & 0.00 (0.00) & 0.00 (0.00) & 0.00 (0.00) \\
			Metastases & yes &  \bf{1.35 (0.28)} &  \bf{0.57 (0.11)} &  \bf{ 0.94 (0.18)} &  \bf{0.89 (0.12)} &  \bf{-0.19 (0.08)} & 0.00 (0.00) & -0.05 (0.05) & 0.00 (0.00) \\
			& unknown &   \bf{0.83 (0.30)} & 0.17 (0.19) &  \bf{0.43 (0.13)} &  \bf{0.53 (0.13)} & -0.14 (0.09) & 0.01 (0.07) & 0.00 (0.00) & 0.00 (0.00) \\
			Sodium level & $< 136\ mmol/l$ &  \bf{0.33 (0.14)} &  \bf{0.27 (0.09)} & \bf{ 0.31 (0.14)} &  0.14 (0.08) & -0.01 (0.05) & 0.00 (0.00) & 0.00 (0.00) & 0.00 (0.00) \\
			& missing &  -0.77 (0.45) & 0.00 (0.00) & 0.00 (0.00) & 0.00 (0.00) &  \bf{0.32 (0.16)} & 0.02 (0.09) & 0.04 (0.09) & 0.00 (0.00) \\
			Albumen level & $< 35\ g/l$ &  \bf{0.65 (0.16)} & \bf{ 0.44 (0.15)} &  \bf{0.50 (0.14)} &  \bf{0.36 (0.09)} & -0.10 (0.06) & -0.04 (0.06) & -0.07 (0.06) & 0.00 (0.00) \\
			& missing &   \bf{0.59 (0.28)} & 0.00 (0.00) & 0.00 (0.00) & 0.00 (0.00) & 0.09 (0.15) &  \bf{0.15 (0.07)} & 0.12 (0.07) & 0.00 (0.00) \\
			\hline
		\end{tabular}}
	\end{adjustbox}
	{\footnotesize\begin{tablenotes}
		\item[] Bold indicates statistically significant at the $5\%$ level.
	\end{tablenotes}}
	\end{threeparttable}
\end{sidewaystable}

\begin{sidewaystable}
%\begin{landscape}
%\begin{table}[ht]
	\caption{Coefficients estimates and standard errors for the two tuning parameters setup {(lung cancer dataset)}\label{tab:LC_VSestimates2}} \centering
	%\begin{center}
	\begin{threeparttable}
	\begin{adjustbox}{max width=\linewidth}
		{\footnotesize\begin{tabular}{ll|cccc|cccc}
		\hline
		{} & {} & \multicolumn{4}{c|}{Scale}& \multicolumn{4}{c}{Shape}\\
		\hline
		\multicolumn{2}{c|}{Covariate}& No Penalty& LASSO & SCAD& ALASSO & No Penalty & LASSO & SCAD & ALASSO \\
		Intercept & &  \bf{-3.38 (0.66)} &  \bf{-3.07 (0.25)} &  \bf{-3.62 (0.30)}&  \bf{-3.15 (0.17)} & -0.16 (0.22) &  0.02 (0.06) & 0.05 (0.06) & 0.04 (0.03) \\

		Treatment & surgery &  \bf{-1.69 (0.83)} &  \bf{-1.18 (0.24)} &  \bf{-1.21 (0.25)} &  \bf{-0.98 (0.22)}  & 0.11 (0.21) & 0.00 (0.00) & 0.00 (0.00) & 0.00 (0.00) \\
		& chemotherapy & -0.33 (0.37) & 0.31 (0.21) &  \bf{-0.50 (0.21)} & 0.00 (0.00) & -0.03 (0.15) & 0.00 (0.00) & 0.00 (0.00) & 0.00 (0.00) \\

		& radiotherapy &  \bf{-0.85 (0.21)} &  \bf{-0.37 (0.18)} &  \bf{-0.56 (0.19)} &  \bf{-0.21 (0.10)} &   \bf{0.22 (0.08)} & 0.05 (0.07) & 0.11 (0.07) & 0.00 (0.00) \\

		& chemo. \& radio. &  \bf{-3.83 (0.98)} &  \bf{-0.74 (0.23)} &  \bf{-1.93 (0.81)} &  \bf{-0.63 (0.22)} &   \bf{0.77 (0.20)} & 0.00 (0.00) & 0.33 (0.22) & 0.00 (0.00) \\
		Age group & $50-$ &  \bf{-0.90 (0.43)} & 0.00 (0.00) & 0.00 (0.00) & 0.00 (0.00) &  \bf{0.39 (0.16)} & 0.02 (0.06) & 0.02 (0.06) & 0.00 (0.00) \\

		& $60-$ &  \bf{-0.94 (0.39)} & 0.00 (0.00) & 0.00 (0.00) & 0.00 (0.00) &  \bf{0.40 (0.15)} & 0.02 (0.04) & 0.02 (0.04) & 0.00 (0.00) \\

		& $70-$ & -0.77 (0.39) & 0.00 (0.00) & 0.00 (0.00) & 0.00 (0.00) &  \bf{0.31 (0.15)} & 0.00 (0.00) & 0.00 (0.00) & 0.00 (0.00) \\

		& $> 80$ & -0.78 (0.42) & 0.00 (0.00) & 0.00 (0.00) & 0.00 (0.00) & 0.31 (0.17) & -0.01 (0.05) & -0.01 (0.05) & 0.00 (0.00)\\

		WHO status & light work & -0.02 (0.45) &-0.20 (0.26) & 0.00 (0.00) & 0.00 (0.00) & 0.02 (0.12) & 0.08 (0.08) & 0.09 (0.05) & 0.00 (0.00) \\
		& unable to work & 0.84 (0.43) &  \bf{0.40 (0.17)} &  \bf{0.66 (0.15)} & \bf{0.44 (0.10)} & -0.10 (0.13) & 0.00 (0.00) & 0.00 (0.00) & 0.00 (0.00) \\

		& $> 50\%$ walking &  \bf{1.31 (0.44)} &  \bf{0.90 (0.19)} &  \bf{1.14 (0.16)} &  \bf{0.97 (0.11)} & -0.13 (0.14) & 0.00 (0.00) & 0.00 (0.00) & 0.00 (0.00) \\

		& bed/chair bound &  \bf{1.78 (0.50)} &  \bf{1.50 (0.29)} &  \bf{1.79 (0.26)} &  \bf{1.54 (0.25)}  & -0.03 (0.20) & 0.00 (0.00) & 0.00 (0.00) & 0.00 (0.00) \\

		Sex & male & 0.03 (0.14) & 0.00 (0.00) & 0.00 (0.00) & 0.00 (0.00) & -0.03 (0.05) & 0.00 (0.00) & 0.00 (0.00) & 0.00 (0.00) \\

		Smoking status & current smoker & 0.10 (0.22) & 0.12 (0.08) & 0.22 (0.17) & 0.00 (0.00) & 0.15 (0.08) & 0.00 (0.00) &0.00 (0.00) & 0.00 (0.00) \\
		& ex-smoker &  -0.05 (0.23) & 0.00 (0.00) & 0.06 (0.26) & 0.00 (0.00) & 0.17 (0.09) & 0.00 (0.00) & 0.00 (0.00) & 0.00 (0.00) \\

		& missing & 0.29 (0.40) & 0.00 (0.00) & 0.00 (0.00) & 0.00 (0.00)& 0.00 (0.19) & 0.00 (0.00) & 0.00 (0.00) & 0.00 (0.00) \\

		Cell type & small cell &  \bf{0.83 (0.26)} &  \bf{0.52 (0.16)} &  \bf{0.72 (0.16)} & \bf{0.43 (0.13)} & -0.05 (0.10) & 0.00 (0.00) & 0.00 (0.00) & 0.00 (0.00) \\
		& adenocarcinoma & 0.28 (0.28) & 0.14 (0.14) &  \bf{0.30 (0.14)} & 0.00 (0.00) & 0.03 (0.10) & 0.00 (0.00) & 0.00 (0.00) & 0.00 (0.00) \\

		& other & 0.32 (0.20) & 0.13 (0.10) & 0.24 (0.16) & 0.09 (0.09) & -0.04 (0.07) & 0.00 (0.00) & -0.01 (0.06) & 0.00 (0.00) \\

		Metastases & yes &  \bf{1.35 (0.28)} &  \bf{0.67 (0.12)} &  \bf{0.77 (0.12)} &  \bf{0.84 (0.12)} &  \bf{-0.19 (0.08)} & 0.00 (0.00) & 0.00 (0.00) & 0.00 (0.00) \\
		& unknown &  \bf{0.83 (0.30)} &  \bf{0.26 (0.13)} &  \bf{0.35 (0.13)} &  \bf{0.41 (0.13)} & -0.14 (0.09) & 0.00 (0.00) & 0.00 (0.00) & 0.00 (0.00) \\
		Sodium level & $< 136\ mmol/l$ &  \bf{0.33 (0.14)} &  \bf{0.30 (0.08)} &  \bf{0.32 (0.08)} &  \bf{0.24 (0.08)}& -0.01 (0.05) & 0.00 (0.00) & 0.00 (0.00) & 0.00 (0.00) \\
		& missing &  -0.77 (0.45) & 0.00 (0.00) & 0.00 (0.00) & 0.00 (0.00) &  \bf{0.32 (0.16)} & 0.00 (0.00) & 0.00 (0.00) & 0.00 (0.00) \\
		Albumen level & $< 35\ g/l$ &  \bf{0.65 (0.16)} &  \bf{0.46 (0.15)} &  \bf{0.59 (0.15)} &  \bf{0.37 (0.09)} & -0.10 (0.06) & -0.04 (0.06) & -0.08 (0.06) & 0.00 (0.00)\\
		& missing &  \bf{0.59 (0.28)} &  \bf{0.36 (0.14)} &  \bf{0.43 (0.14)} &  \bf{0.27 (0.14)} & -0.06 (0.12) & 0.00 (0.00) & 0.00 (0.00) & 0.00 (0.00) \\
		\hline
	\end{tabular}}
\end{adjustbox}
{\footnotesize\begin{tablenotes}
	\item[] Bold indicates statistically significant at the $5\%$ level.
\end{tablenotes}}
\end{threeparttable}
\end{sidewaystable}

\end{document}